\documentclass[sn-mathphys,Numbered]{sn-jnl}

\usepackage{graphicx}%
\usepackage{multirow}%
\usepackage{amsmath,amssymb,amsfonts}%
\usepackage{amsthm}%
\usepackage{mathrsfs}%
\usepackage{dsfont}
\usepackage[title]{appendix}%
\usepackage{xcolor}%
\usepackage{textcomp}%
\usepackage{manyfoot}%
\usepackage{booktabs}%
\usepackage{algorithm}%
\usepackage{algorithmicx}%
\usepackage{algpseudocode}%
\usepackage{listings}%

\usepackage{soul}

\DeclareMathOperator{\Tr}{Tr}
\begin{document}

\title[Article Title]{A linear photonic swap test circuit for quantum kernel estimation}


\author*[1]{\fnm{Alessio} \sur{Baldazzi}}\email{alessio.baldazzi@unitn.it}
\equalcont{These authors contributed equally to this work.}

\author[1]{\fnm{Nicolò} \sur{Leone}}\email{nicolo.leone@unitn.it}
\equalcont{These authors contributed equally to this work.}

\author[1]{\fnm{Matteo} \sur{Sanna}}\email{matteo.sanna@unitn.it}
\equalcont{These authors contributed equally to this work.}

\author[1]{\fnm{Stefano} \sur{Azzini}}\email{stefano.azzini@unitn.it}

\author[1]{\fnm{Lorenzo} \sur{Pavesi}}\email{lorenzo.pavesi@unitn.it}

\affil[1]{\orgdiv{Department of Physics}, \orgname{University of Trento}, \orgaddress{\street{Via Sommarive 14}, \city{Trento}, \postcode{38123}, \country{Italy}}}


\abstract{
Among supervised learning models, Support Vector Machine stands out as one of the most robust and efficient models for classifying data clusters. At the core of this method, a kernel function is employed to calculate the distance between different elements of the dataset, allowing for their classification. Since every kernel function can be expressed as a scalar product, we can estimate it using Quantum Mechanics, where probability amplitudes and scalar products are fundamental objects. The swap test, indeed, is a quantum algorithm capable of computing the scalar product of two arbitrary wavefunctions, potentially enabling a quantum speed-up. Here, we present an integrated photonic circuit designed to implement the swap test algorithm. Our approach relies solely on linear optical integrated components and qudits, represented by single photons from an attenuated laser beam propagating through a set of waveguides. By utilizing 2$^3$ spatial degrees of freedom for the qudits, we can configure all the necessary arrangements to set any two-qubits state and perform the swap test. This simplifies the requirements on the circuitry elements and eliminates the need for non-linearity, heralding, or post-selection to achieve multi-qubits gates.
Our photonic swap test circuit successfully encodes two qubits and estimates their scalar product with a measured root mean square error smaller than 0.05. This result paves the way for the development of integrated photonic architectures capable of performing Quantum Machine Learning tasks with robust devices operating at room temperature.

}


\maketitle

\section{Main}
\label{sec:intro}

Quantum Machine Learning (QML) \cite{wittek, rupp2015machine, schuld2015introduction, Biamonte_2017} leverages two rapidly advancing technologies: Quantum Computing (QC) \cite{nielsen_chuang_2010} and Machine Learning (ML) \cite{mitchell, apply}. Within this framework, Quantum Mechanics has been applied to solve classical tasks \cite{Niu2019, Li2021, Moradi2022}, while ML methods have facilitated the understanding of new quantum aspects \cite{rupp2015machinebb, PhysRevLett.120.240501, dong2019machine, gebhart2023learning}.

From this fruitful interaction, a language similarity between Quantum Mechanics and Support Vector Machine (SVM) \cite{cortes1995support} has emerged \cite{wittek, Biamonte_2017, schuld, QML_schuld, Havl_ek_2019}. On one side, wavefunctions can be expressed as a weighted superposition of observables' eigenstates, and on the other side, the classifier parameter in SVMs is determined through a weighted sum of similarities. SVMs, as supervised learning models \cite{schuld2015introduction}, exhibit high robustness in categorizing clusters of data.

These models rely on a kernel function \cite{cristianini2000introduction, patle2013svm}, computed from the scalar product between the data represented in the feature space. This kernel function measures how different test data are from the training dataset. The efficiency of SVMs in classification tasks is largely influenced by the quality of scalar products \cite{kernel, burges1998tutorial}. The computational cost to construct a kernel function scales non-linearly with respect to the dataset dimension, making challenging for classical machines to handle large databases \cite{Convex}.

Quantum Mechanics can offer a linear scaling advantage, where scalar products computation plays a crucial role. By exploiting larger configuration spaces, a quantum speed-up can be achieved through suitable algorithms implemented on quantum hardware. The ultimate goal is a hybrid architecture consisting of an SVM enhanced by quantum hardware that performs quantum algorithms more efficiently, providing input for the ML program \cite{QSVM, cerezo, Biamonte_2017}.

The swap test \cite{swaptest} serves as an example of a quantum algorithm that calculates the scalar product of two generic quantum states by sampling the outcomes of an auxiliary qubit. Thus, by encoding the data in quantum bits, or qubits, the swap test has the potential to achieve a speed-up in estimating the kernel function entries.

Swap test implementations have thus far been realized using superconducting qubits and trapped ions \cite{Cincio_2018, nguyen2021experimental}. These implementations involve low temperatures and complex schemes, utilizing more gates and qubits compared to the circuit expressed in gate-based notation \cite{Cincio_2018}.

Quantum optics and photonics present a promising avenue for swap test implementations, though currently, there are only theoretical proposals for optical setups \cite{Kang2019ImplementationOS}. The primary challenge for the photonic platform lies in realizing the controlled-SWAP (CSWAP) or Fredkin gate \cite{Fredkin1982}. Theoretical proposals and experimental demonstrations of CSWAP gates are grounded in linear optics combined with probabilistic strategies \cite{milburn1989quantum, fiuravsek2006linear, gong2008methods}, cross-Kerr non-linearity \cite{fred_1, Dong:16}, sources of entangled photon pairs \cite{sciadv_patel, PhysRevA.78.032317, Ono2017}, continuous variables version with two-mode squeezed states \cite{Volkoff_2022}, single photons and quantum dots confined in single-sided cavities \cite{Kang2020}, as well as combinations of coherent states, threshold photon detectors, and classical post-processing \cite{PhysRevA.98.062318, PhysRevResearch.3.043035}.
Finally, in \cite{Wang_2021}, a unique approach is taken where a single photon is used to encode three qubits for CSWAP: the control qubit in polarization and the remaining four-dimensional Hilbert space in the photonic orbital angular momentum.

Here, we propose and validate a pioneering photonic integrated circuit (PIC) capable of implementing the swap test between two qubits at room temperature. The encoding of multiple qubits is achieved using single photons propagating in a set of waveguides \cite{PhysRevA.57.R1477}, implementing the equivalence between path-encoded qudits and their corresponding qubits \cite{Wang_2020}. With linear optical devices on-chip and single-photon detectors off-chip, our encoding allows for the representation of generic two-qubits states and the execution of the swap test algorithm using photons from an attenuated laser.

In particular, the preparation and manipulation are accomplished solely through well-established linear optical integrated elements, such as multimode interferometers (MMIs), waveguide crossings (CRs), and thermal phase shifters (PSs) \cite{vivien2013handbook}. Through the detection of single photons containing all the necessary information, we perform the sampling of the swap test's auxiliary qubit with a destructive procedure.

In essence, the PIC is a straightforward, short-depth circuit that utilizes the minimum number of qubits and gates, and it can be directly mapped to the gate-based language. Unlike other photonic proposals, our PIC does not rely on modified versions of the original swap test circuit or non-linear optical phenomena and post-processing.

Our approach serves as a novel building block for optical implementations of Quantum Machine Learning (QML) algorithms, capable of operating at room temperature and characterized by robustness and low technological requirements. By scaling up the computational capability of our quantum machine, it becomes possible to construct an efficient quantum kernel estimator for data cluster classification in the current noisy intermediate-scale quantum era \cite{QSVM, cerezo}.

\begin{figure}[htbp]
\centering
\includegraphics[width=\textwidth]{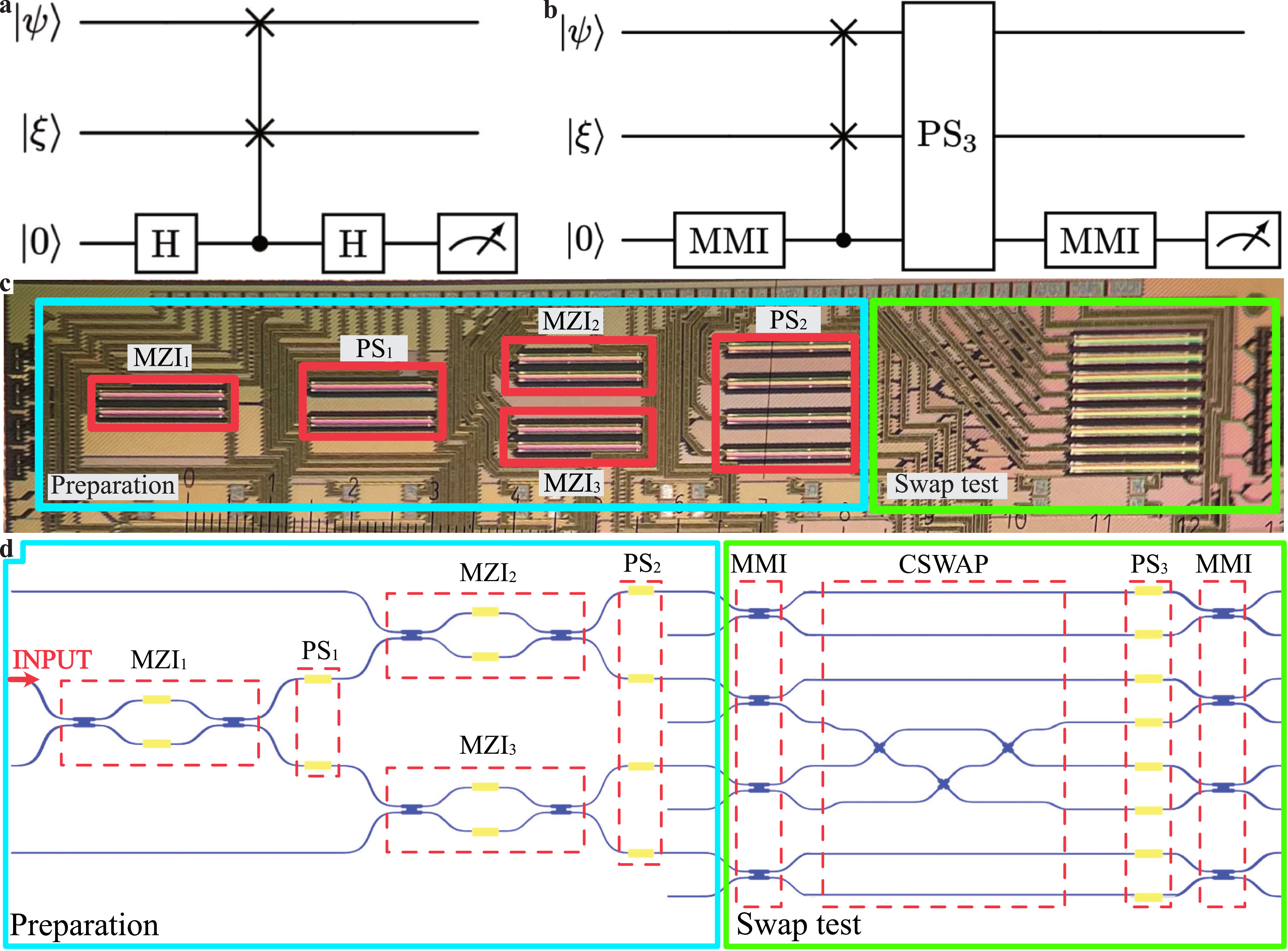}
\caption{\textbf{The PIC for the quantum photonic swap test.} \textbf{a)} Gate representation of the standard swap test algorithm. The circuit starts with the input states $|\psi\rangle$, $|\xi\rangle$ and an auxiliary qubit in the state $|0\rangle$. The H letter stands for Hadamard gate, and the three-qubit gate inserted between the H gates is the controlled-SWAP gate with the third qubit as the control. The algorithm ends with the measurement of the auxiliary qubit. \textbf{b)} Gate representation of our linear photonic swap test circuit. The H gate is implemented through MMIs and the gate ${\rm PS}_3$ represents the action of PSs needed to balance the optical paths. \textbf{c)} Photo of the PIC that implements the swap test circuit. The cyan box (dimension $7\times1$ mm$^2$) contains the preparation stage, where the optical elements $\{\text{MZI}_i\}_{i=1,2,3}$ and $\{\text{PS}_j\}_{j=1,2}$, shown in red boxes, are used to encode the states. The green box contains the swap test circuit (dimension $3\times1$ mm$^2$). \textbf{d)} Schematic representation of the photonic swap test circuit. In the figure, blue color is used for the waveguides, while yellow is for the phase shifters. The preparation stage is shown in the cyan box: this part is used to encode the total input state $|\psi\rangle|\xi\rangle|0\rangle$. The INPUT label indicates the waveguide where light is injected. The green box includes the swap test stage, which is composed of two series of MMIs, our CSWAP implementation and the phase shifting elements $\text{PS}_3$.}
\label{photo}
\end{figure}

\begin{figure}[htbp]
\centering
\includegraphics[width=\textwidth]{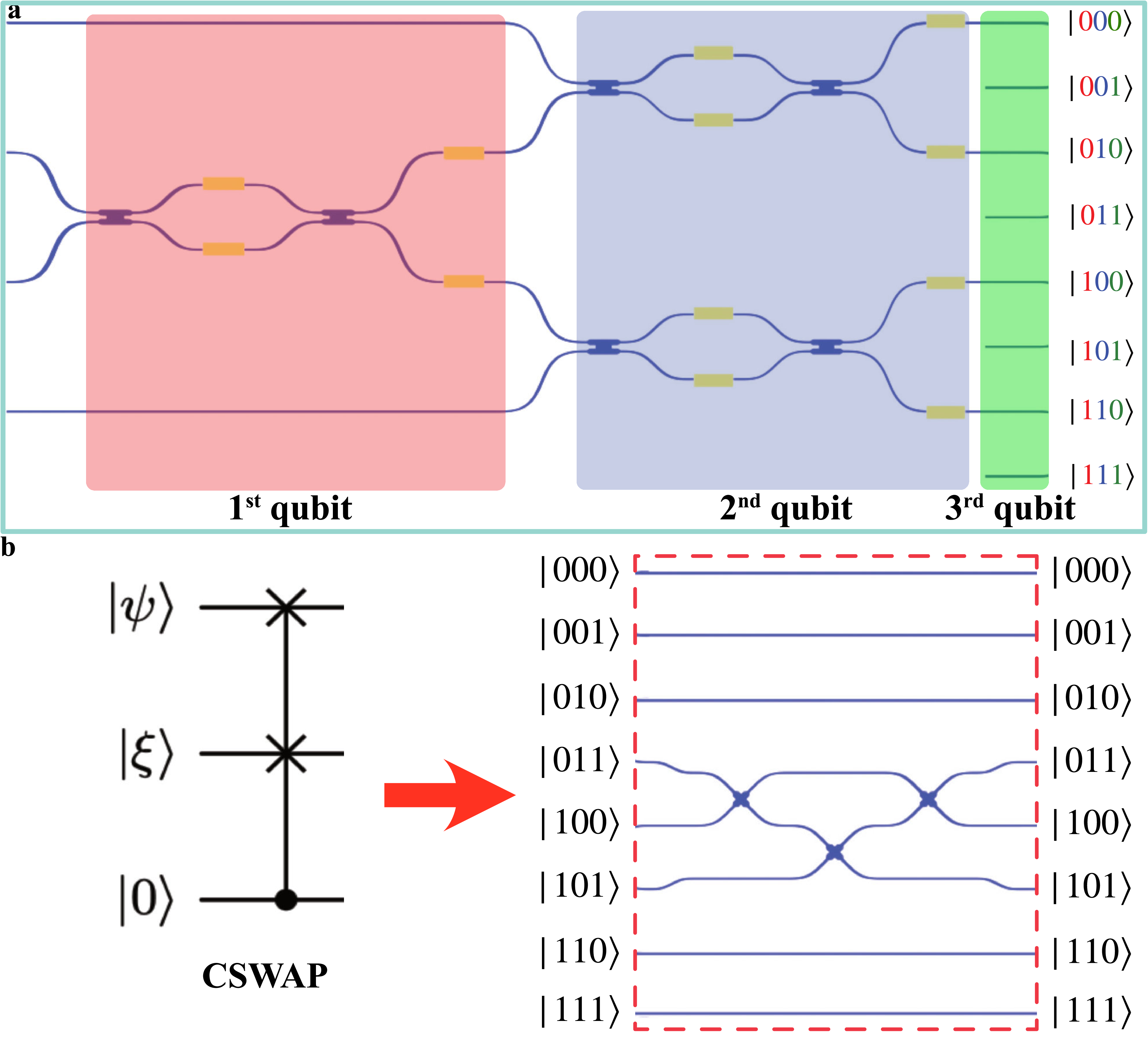}
\caption{\textbf{Details of the quantum photonic swap test circuit.} \textbf{a)} Detail of the preparation stage (cyan box). 
In the red square are indicated the MZI and the PSs used to set the state $|\psi\rangle$. In blue we highlight the MZIs and the PSs used to encode the state $|\xi\rangle$. The MZIs are composed of two MMIs and two PSs, represented by yellow boxes, in the internal branches. Eventually, in green it is shown the initialization of the third qubit, the ancilla, to the state $|0\rangle$. This is achieved by connecting the even-numbered waveguides to the previous circuit and leaving the others unconnected. \textbf{b)} Gate representation of CSWAP operation (left) and our implementation based on path encoding (right). Note that the optical elements reported in the figure are waveguide crossings.}
\label{circuit}
\end{figure}

\section{Linear photonic circuit implementing the swap test algorithm}
\label{sec:photo_design}

The swap test \cite{swaptest} is a quantum algorithm designed to assess the overlap between two input quantum states (more details are provided in Supplementary Information Section \ref{app:preliminaries}). The degree of overlap is quantified by the squared scalar product of the inputs, indicating the dissimilarity between the two states \cite{Li2021}.

In Fig. \ref{photo}(a), the quantum gate notation for the swap test is illustrated. The algorithm begins with two qubits, $|\psi\rangle$ and $|\xi\rangle$, for which we seek to determine the scalar product. An auxiliary qubit, referred to as the ancilla, is initialized to the state $|0\rangle$. The first Hadamard (H) gate generates a superposition of $|0\rangle$ and $|1\rangle$ for the ancilla. Subsequently, a CSWAP gate is applied to the overall state, using the auxiliary qubit as the control qubit. The CSWAP outcome involves the swapping of the two target qubits if the control is in the state $|1\rangle$ and no swapping otherwise. Following this operation, the H gate is applied again to the auxiliary qubit.

Finally, the ancilla is measured, and the probabilities of the outcomes, denoted as ${\mathbb{P}(x)}_{x=0,1}$, are algebraically related to the squared scalar product of the two input states. Specifically, it holds:

\begin{equation}
    |\langle \psi| \xi \rangle|^2 = 2 \,\mathbb{P}(0) - 1 
    =
    1- 2 \,\mathbb{P}(1)
    \,.
    \label{prob_result_swap_stand}
\end{equation}
Similar to other quantum algorithms, such as the one proposed in \cite{Peruzzo_2014}, a computing problem is translated into a sampling problem. Therefore, once sufficient statistics on the ancilla's outcomes are collected, the overlap of the two qubits can be efficiently calculated.

We conduct a direct translation of the gates in the standard scheme, depicted in Fig. \ref{photo}(a), into the corresponding photonic operations, as represented in Fig. \ref{photo}(b). Fig. \ref{photo}(c-d) displays a photograph and the schematics of the Photonic Integrated Circuit (PIC), respectively. The PIC consists of two main parts: the preparation stage (cyan box in Fig. \ref{photo}(c-d)), where the state $|\psi\rangle \otimes |\xi\rangle \otimes |0\rangle$ is set, and the swap test stage (green box in Fig. \ref{photo}(c-d)), where our swap test algorithm, illustrated in Fig. \ref{photo}(b), is executed.

To encode states in the PIC, we employ path-encoded single photons, where the state is determined by the waveguide through which a single photon travels \cite{adami1998quantum, Knill2001, Kok_2007}. For instance, in a setup with two waveguides, the state $|0\rangle$ corresponds to a photon in the upper waveguide, while the state $|1\rangle$ corresponds to a photon in the lower waveguide. Consequently, when we have $2^n$ waveguides and one photon, we have a qudit, and we assign the state $|i\rangle$, with $i \in {0, \dots, 2^n-1}$, to the photon in the $i$-th waveguide \cite{Wang_2020}.

In our proof-of-concept experiment, we utilize photons from an attenuated laser to prepare the two qubits $|\psi\rangle$ and $|\xi\rangle$ as inputs for the swap test, and one qubit is set to $|0\rangle$ for the ancilla. Therefore, eight paths, i.e. $2^3$, are necessary. The encoding is achieved by simply representing the waveguide number and converting it to binary numbers:
\begin{equation}
\begin{split}
    &|0\rangle \to |000\rangle \equiv |0\rangle\otimes|0\rangle\otimes|0\rangle\,, \qquad
     |1\rangle \to |001\rangle \equiv |0\rangle\otimes|0\rangle\otimes|1\rangle\,, \\
     &|2\rangle \to |010\rangle \equiv |0\rangle\otimes|1\rangle\otimes|0\rangle\,, \qquad
     |3\rangle \to |011\rangle \equiv |0\rangle\otimes|1\rangle\otimes|1\rangle\,, \\
      &|4\rangle \to |100\rangle \equiv |1\rangle\otimes|0\rangle\otimes|0\rangle\,, \qquad
     |5\rangle \to |101\rangle \equiv |1\rangle\otimes|0\rangle\otimes|1\rangle\,, \\
      &|6\rangle \to |110\rangle \equiv |1\rangle\otimes|1\rangle\otimes|0\rangle\,, \qquad
     |7\rangle \to |111\rangle \equiv |1\rangle\otimes|1\rangle\otimes|1\rangle\,, 
     \label{eq:queight}
\end{split}
\end{equation}
where the state on the left-hand side of the arrow represents the qudits state, while the first two positions on the right-hand side of the arrows belong to the qubits, whose scalar product we are interested in, and the third position pertains to the ancilla.
This means that when a photon is in the top waveguide, or equivalently, the qudit is in the state $|0\rangle$, the corresponding prepared three-qubit state is $|000\rangle$. The same principle applies to the remaining states, as illustrated in the right part of Fig. \ref{circuit}(a).

The preparation stage consists of three Mach-Zehnder Interferometers (MZIs) and six Phase Shifters (PSs), indicated by the red boxes in Fig. \ref{photo}(c-d).

The encoding of the input qubits $|\psi\rangle$ and $|\xi\rangle$ involves three steps, illustrated by the red, blue, and green rectangles in Fig. \ref{circuit}(a). The first step, comprising one MZI and two PSs ($\text{MZI}_1$ and $\text{PS}_1$ in Fig. \ref{photo}(c-d)), establishes the state of the first qubit by imposing phases $\left(\Delta\theta_1, \Delta\phi_1\right)$. The second step, involving the second and third MZIs along with the last four PSs ($\text{MZI}_2$, $\text{MZI}_3$, and $\text{PS}_2$ in Fig. \ref{photo}(c)), prepares the second qubit by setting phases $\left(\Delta\theta_2, \Delta\phi_2\right)$.
Modulo a global phase, the prepared state reads:\begin{equation}
| \psi \rangle \otimes | \xi \rangle =
\big(\sin \Delta\theta_1|0\rangle + {\rm e}^{-{\rm i} \Delta\phi_1}\cos \Delta\theta_1|1\rangle \big)
\!\otimes\!
\big(\sin \Delta\theta_2|0\rangle + {\rm e}^{-{\rm i} \Delta\phi_2}\cos \Delta\theta_2|1\rangle \big)\, .
\label{eq:twoqubit_prep}
\end{equation}
This expression is the generic form of a separable state of two qubits. In the third and final step of the preparation stage (green rectangles in Fig. \ref{circuit}(a)), the outputs of $\text{PS}_2$ are connected to the even-numbered waveguides of the $2^3$-qudit. Consequently, the ancilla is passively initialized to the state $|0\rangle$, and the prepared total state reads:
\begin{equation}
\label{state_tot}
   |\Psi\rangle= | \psi \rangle \otimes | \xi \rangle \otimes |0\rangle \,.
\end{equation}
Here, $| \psi \rangle \otimes | \xi \rangle$ is defined by equation \eqref{eq:twoqubit_prep}, and the numbering logic for the states is provided in equation \eqref{eq:queight} and depicted on the right side of Fig. \ref{circuit}(a).

After setting the state, it serves as the input for the next stage. This is the swap test stage, comprising four layers, as illustrated in Fig. \ref{photo}(d) and enclosed within the green box in Fig. \ref{photo}(c). The corresponding evolution operator is expressed as (refer to Supplementary Information Section \ref{app:theory} for details):
\begin{align}
\label{U_swap}
U_{\rm swaptest}=\left(\mathds{1} \otimes \mathds{1} \otimes U_{\rm MMI}\right) \cdot 
U_{\text{PS}_3}
\cdot U_{\rm CSWAP} \cdot
\left(\mathds{1} \otimes \mathds{1} \otimes U_{\rm MMI}\right) \,.
\end{align}
The transformation $\left(\mathds{1} \otimes \mathds{1} \otimes U_{\rm MMI}\right)$ corresponds to a layer of four parallel beam splitters implemented using MMIs between every pair of waveguides. $U_{\text{PS}3}$ represents the layer of thermal PSs, and $U_{\rm CSWAP}$ describes the action of the CRs network implementing the CSWAP gate.

As illustrated in Fig. \ref{circuit}(b), the CSWAP gate consists of three CRs realizing the swapping of $|011\rangle$ and $|101\rangle$, modulo overall phases. Remarkably, this operation occurs in a passive circuit, contrasting with the complexity of other photonic approaches reported to date \cite{fred_1, Dong:16, Kang2019ImplementationOS}.
The presence of PS$_3$ in the PIC is necessary to correct spurious phases between the optical paths; ideally, our swap test stage could be achieved in a completely passive manner.

Finally, once the transformation \eqref{U_swap} is applied to the state \eqref{state_tot}, the ancilla is measured. The probability $\{\mathbb{P}(x)\}_{x=0,1}$ of observing the ancilla in the state $|x\rangle$ is given by:
\begin{equation}
\mathbb{P}(x) 
=\Tr[U_{\rm swaptest} |\Psi\rangle\langle\Psi| U_{\rm swaptest}^\dag \!\cdot \left(\mathds{1} \otimes \mathds{1} \otimes P_x \right)]=\frac{1}{2} \left( 1 - (-)^x |\langle \psi| \xi \rangle|^2 \right) \,,
\label{eq:ourswap_res}
\end{equation}
where $P_x$ is the projector onto the state $|x\rangle$ for the ancilla, and $|\psi\rangle$ and $|\xi\rangle$ are defined in equation \eqref{eq:twoqubit_prep}. The theoretical result for the scalar product $|\langle \psi| \xi \rangle|^2$, recalling equation \eqref{eq:twoqubit_prep}, is expressed as:
\begin{equation}
    |\langle \psi| \xi \rangle|^2= \cos^2\!\left( \Delta\theta_1\!-\!\Delta\theta_2\right) \cos^2\!\!\left( \!\frac{\Delta\phi_1\!-\!\Delta\phi_2}{2}\!\right)
    \!+
    \cos^2\!\left( \Delta\theta_1\!+\!\Delta\theta_2\right) \sin^2\!\!\left( \!\frac{\Delta\phi_1\!-\!\Delta\phi_2}{2}\!\right)
    .
    \label{eq:sca_pro_formula}
\end{equation}
Note that our result, as shown in equation \eqref{eq:ourswap_res}, is equal to that of the standard swap test algorithm, as reported in equation \eqref{prob_result_swap_stand}, except for the exchange of $\mathbb{P}(0)$ and $\mathbb{P}(1)$. This difference arises from the implementation of the H gate using the MMI transformation.

Experimentally, the probabilities $\{\mathbb{P}(x)\}_{x=0,1}$ are estimated by sampling the counts at the even and odd outputs of the ancilla:
\begin{equation}
\mathbb{P}(x)=\frac{N_x}{N_0+N_1},
\end{equation}
where $N_0$ and $N_1$ are the photon counts coming from even and odd waveguides, respectively. The counts are measured by tuning the phases of the operation $U_{\text{PS}_3}$ in the swap test stage and sampling only the even outputs with four detectors and twice the acquisition time. Since the total state \eqref{state_tot} is written in single photons at each run, the detection stage makes the implemented scheme destructive, even though it is based on the non-destructive swap test scheme.

It is crucial to emphasize that the adoption of path encoding and qudits is the cornerstone of our PIC. This choice simplifies the circuit, comprising only linear optical transformations, and, as a result, allows the use of an attenuated laser source. This is possible because such operations act linearly at the single-photon level (refer to Supplementary Information Section \ref{app:att_source}).

\section{Results}
\label{sec:results}

Our PIC is validated using various sets of states, and the experimental results are compared with theoretical outcomes, as given by equation \eqref{eq:sca_pro_formula}. Additionally, we model the real characteristics of our PIC, considering factors such as the imbalance of MMIs, the insertion loss of CRs, and calibration errors of the phases of the MZIs (refer to Supplementary Information Section \ref{app:ni_model}). We calculate a $2\sigma$ confidence interval for the outcomes.

\begin{figure}[]
\centering
  \includegraphics[width=\textwidth]{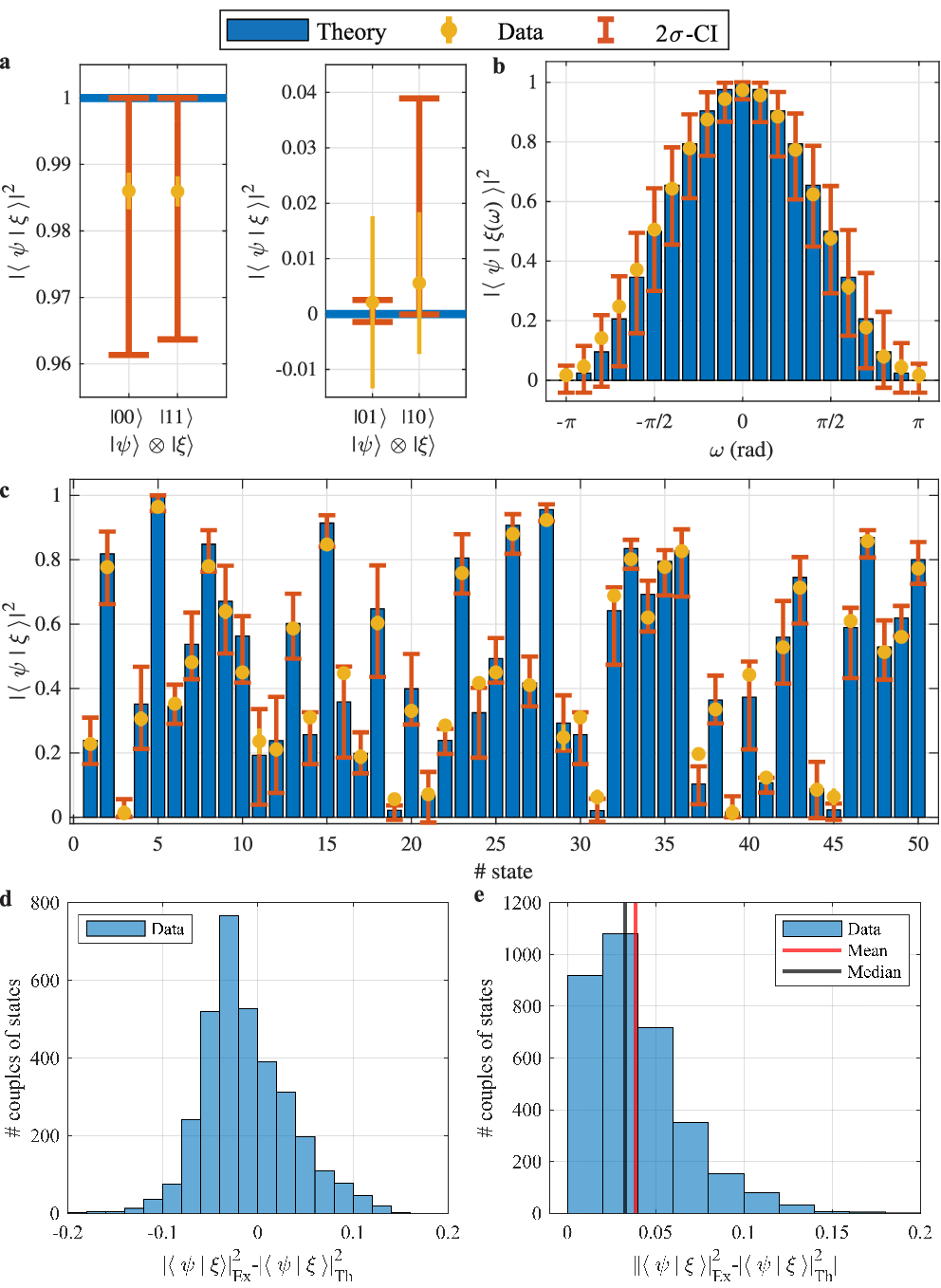}
\caption{\textbf{Results on different ensembles of pairs of states.} 
\textbf{a)} Results for the computational basis states in the parallel ($|00\rangle$ and $|11\rangle$) and orthogonal cases ($|01\rangle$ and $|10\rangle$). The theoretical squared scalar products are indicated in blue. The measured data with their error bars are reported in gold. Errors are obtained through repeated measurements. The $2\sigma$ confidence interval ($2\sigma$-CI) is reported in orange.
\textbf{b)}
Results for the family of states' pairs of equation \eqref{eq:parallel}. The theoretical squared scalar products are shown in blue, while the data collected from repeated measurements are reported in gold together with their error bars. The $2\sigma$ confidence interval is reported in orange.
\textbf{c)} Subset of 50 pairs of states randomly chosen between the used 3342 pairs. The light blue bar represents the expected theoretical value for the squared scalar product, the red interval indicates the $2\sigma$ confidence interval as provided by the model and, lastly, the experimental data, with their error bars are reported as gold dots. \textbf{d)} Histogram of the experiment-theory distance $|\langle \psi | \xi \rangle|^2_{\rm Ex}-|\langle \psi | \xi \rangle|^2_{\rm Th}$. 
\textbf{e)} Histogram of the absolute distance $||\langle \psi | \xi\rangle|^2_{\rm Ex}-|\langle \psi | \xi\rangle|^2_{\rm Th}|$. 
Median (black line) and mean (red line) are also shown. The histogram's bin has width $0.02$ and is centered around multiple values of $0.01$.
}
 
\label{fig:result_hist}
\end{figure}

A preliminary validation is conducted on the set of states $\{|0\rangle\otimes|0\rangle, |0\rangle\otimes|1\rangle, |1\rangle\otimes|0\rangle, |1\rangle\otimes|1\rangle\}$, representing the computational basis states. In Fig. \ref{fig:result_hist}(a), theoretical values of the corresponding squared scalar products are depicted in blue, while experimental measurements, along with error bars obtained from multiple runs on our PIC, are shown in gold. The results align well with theory for pairs of orthogonal states $\{|0\rangle\otimes|1\rangle, |1\rangle\otimes|0\rangle \}$ and exhibit a slight deviation for pairs of parallel states $\{|0\rangle\otimes|0\rangle, |1\rangle\otimes|1\rangle \}$.
This deviation is attributed to actual fabrication errors in our PIC, as the data fall within the $2\sigma$ confidence interval computed using the real characteristic model (red error bars in Fig. \ref{fig:result_hist}(a)). Thus, our PIC effectively estimates the squared scalar product of the computational basis states.

In the second step of our validation process, we consider the following family of states:
\begin{equation}
|\psi\rangle = \frac{1}{\sqrt{2}}\big(|0\rangle + |1\rangle \big)
\quad , \quad
|\xi(\omega)\rangle = \frac{1}{\sqrt{2}}\big(|0\rangle + {\rm e}^{{\rm i} \omega}|1\rangle \big)\, ,
\label{eq:parallel}
\end{equation}
where $\omega$ is a free parameter in the state $|\xi\rangle$. By varying only the current of the PS determining $\omega$ (the second PS from the top in $\text{PS}_2$, see Fig. \ref{photo}(d) and \ref{circuit}(a)), we study the behavior of the PIC. This allows us to explore situations ranging from the initial pair of parallel states ($\omega=0$) to the orthogonal case ($\omega=\pm\pi$).

The obtained data are presented in Fig. \ref{fig:result_hist}(b) as gold dots with their error bars. The experimental data closely reproduce the theoretical results (blue squares) and exhibit a root mean square error (RMSE) of $0.026$. Moreover, all the data fall within the $2\sigma$ confidence interval obtained from our realistic model. Thus, we confirm that our PIC can effectively estimate the squared scalar product of pairs of states from the family of states given in equation \eqref{eq:parallel}.

Finally, the validation process of our PIC's functionality concludes by randomly selecting 3342 pairs of states $|\psi\rangle$ and $|\xi\rangle$ and calculating their squared scalar product using the PIC. Fig. \ref{fig:result_hist}(c) shows a subset of 50 pairs of states randomly selected from the 3342 pairs.

The PIC accurately computes the squared scalar products for most pairs of random states, and deviations from theory typically fall within the confidence interval of our model. The histogram in Fig. \ref{fig:result_hist}(d) illustrates the distance of the squared scalar product obtained by our PIC from the theoretical value, $|\langle \psi | \xi \rangle|^2_{\text{Ex}}-|\langle \psi | \xi \rangle|^2_{\text{Th}}$, for all 3342 randomly selected pairs of states. The histogram's bin width is $0.02$ and is centered around multiple values of $0.01$. All the distance values between theory and experiments lie between $-0.191$ and $0.144$, with a mean value of $-0.012\pm0.001$.

The asymmetry of the histogram with respect to zero is primarily due to the PIC's characteristics, which tend to decrease the calculated squared scalar product, especially for scalar products near 1. Fig. \ref{fig:result_hist}(e) displays the histogram of the absolute value $||\langle \psi | \xi \rangle|^2_{\text{Ex}}-|\langle \psi | \xi \rangle|^2_{\text{Th}}|$, highlighting both the mean and the median of the distribution. The mean distance is $0.039\pm0.001$, and the squared scalar product of half of the considered states is estimated with an error lower than $0.033$. The observed RMSE slightly increases to $0.048$ compared to the previous test on a much smaller set of states.
Our PIC provides squared scalar products within the $2\sigma$ confidence interval for approximately 88 \% of pairs of states within the random ensemble. Our realistic model captures the systematic non-zero mean value of the distance $|\langle \psi | \xi \rangle|^2_{\text{Ex}}-|\langle \psi | \xi \rangle|^2_{\text{Th}}$ and describes the spread of the data around the corresponding expected theoretical values. Results outside the $2\sigma$ confidence interval are attributed to thermal crosstalk among the MZIs and PSs present in the preparation stage for high applied electrical power, not described by our modeling.

These measurements validate the performance of our PIC in estimating the squared scalar product over a large ensemble of states' pairs. Key figures of merit include a mean absolute error of $0.039$ and a RMSE of $0.048$ with respect to the theory. Moreover, the employed model for PIC's non-idealities not only aids in our understanding of the real PIC's functioning, valuable for realizing improved versions, but also facilitates establishing confidence intervals for the PIC's outputs in kernel estimation.

\section{Discussion}
\label{sec:conclusion}

\begin{table}[!h]
    \centering
    \begin{tabular}{c||c|c|c|c|c}
       & & \multicolumn{3}{|c|}{root mean square error} & \\
          platform   & \# qubits & swap test  & ABA & BBA & Ref. \\ \hline
          \hline
        integrated photonic  & 3    & 0.05 & - & - &  our work\\
         superconducting IBM & 5  & 0.31 &  0.11  & 0.12 &\cite{Cincio_2018} \\
         superconducting Rigetti & 19    & 0.54 & 0.43 & 0.16 & \cite{Cincio_2018} \\
         trapped ion & 5 & not reported & - & - & \cite{nguyen2021experimental}
    \end{tabular}
    \caption{\textbf{Comparison of performances of quantum circuits for states overlap estimation}.
    Our results are compared with those obtained from quantum circuits in various quantum platforms, considering the number of qubits and root mean square error (RMSE) across different algorithms. The symbol '-' is used to denote cases where no experiments have been carried out. }
    \label{tab_comp}
\end{table}

In this study, we present the design and performance of the first photonic integrated circuit (PIC) capable of estimating the scalar product between any two qubits by executing the swap test algorithm. The key advantages of our circuit lie in its simplicity and robustness. The PIC receives photons from an attenuated laser, comprising only well-established integrated devices such as multimode interferometers (MMIs), waveguide crossings (CRs), and phase shifters (PSs). Moreover, it utilizes the minimum number of qubits and gates, maintaining a one-to-one correspondence with the gate-based swap test. Therefore, our implementation represents a resource-efficient realization of the swap test algorithm.

As a benchmark, we provide in Table \ref{tab_comp} the only available data for states overlap estimation acquired on superconducting quantum circuits via swap test or other algorithms, such as the ABA and BBA \cite{Cincio_2018}, involving a higher number of qubits and resources. Notably, other experimental proofs, such as those with trapped ions \cite{linke2018measuring}, only demonstrate qualitative results, hindering a quantitative comparison. Table \ref{tab_comp} demonstrates that our room temperature PIC estimates the squared scalar product with a root mean square error twice lower than the best results reported in the literature for the superconducting platform that requires cryogenic temperatures.

In addition, our proof-of-concept PIC exhibits a mean power consumption per single evaluation of $1.2$ W, with specific allocations of $0.25$ W for the preparation stage, $0.3$ W and $0.4$ W for the swap test stage aimed at sampling the ancilla states $|0\rangle$ and $|1\rangle$, respectively. The sampling time lasts $60$ ms, repeated ten times only to acquire more statistics. These values can be further optimized through design and fabrication improvements, sacrificing statistics, or using more advanced single-photon detectors with shorter dead times.

Our approach can be easily scaled up to operate with large data clusters using arrays of swap test PICs. In \cite{pastorello2023}, the authors demonstrate the construction of a two-layer feedforward neural network using swap test modules, which can be implemented by parallelizing multiple copies of our photonic scheme. Another approach to scaling up leverages the intrinsic scalability of integrated photonic platforms and technological advancements in detector integration \cite{bernard2021top}. Increasing the number of waveguides where photons propagate allows the encoding of more qubits. For instance, the preparation and sampling of $n$-qubits require $\mathcal{O}\left(2^{n}\right)$ paths, MZIs, and detectors. For an 8-qubit state this accounts to 255 MZIs and 256 waveguides, which is well within the capability of silicon photonics \cite{bao2023very}.
The number of detectors can be reduced by multiplexing. One example involves two detectors and coherent multiplexing of all the output waveguides corresponding to the ancillary states $|0\rangle$ and $|1\rangle$. This can be achieved using two large-area single-photon detectors, such as an array of single-photon avalanche diodes (SPADs), placed in front of the output waveguides. Another strategy involves just one detector and an optical switch to recursively reconstruct the probability distribution of outcomes at the cost of increased acquisition time.

Classical methods to estimate kernel entries are physically limited by the required resources for huge numbers of data. Instead, quantum specific-purpose machines have the potential to improve the computational speed and the amount of resources.
The presented results and future outlines demonstrate that our circuit is a promising candidate as a fundamental building block in Quantum Machine Learning applications, estimating kernel function entries over a large dataset. Therefore, a classical Support Vector Machine (SVM) can be enhanced by a resource-efficient quantum kernel estimator comprising scaled-up versions or paralleled compositions of our PIC.

\section{Methods}
The photonic chip has been fabricated using a photolithographic process on an MPW run through a commercially available service offered by Ligentec SA foundry. The waveguide core ($150$ nm-thick and $550$ nm-wide) is made of Silicon Nitride ($\text{Si}_3\text{N}_4$). Waveguides' width has been chosen to have monomodal propagation at the working wavelength of $750$ nm and for the chosen polarization (transverse electric, TE). The $750$ nm wavelength has been determined by the potential of direct SPAD integration in silicon photonics \cite{bernard2021top}. Extension of the PIC design to other wavelengths is straightforward.

All swap test experiments have been carried out using an attenuated Ti:Sapphire laser tuned at $750$ nm as the light source. The light coming from the laser has been attenuated using a variable optical attenuator, and its polarization is controlled and set to be TE at the chip input. In Supplementary Information Section \ref{app:att_source}, a discussion about the use of an attenuated source instead of a true single-photon source is reported.

Optical coupling at the chip's input has been performed using a tapered lensed fiber, while output coupling has been done using a standard fiber array. A power supply (Qontrol systems - BP8 device model) provides the currents for the MZIs and PSs of the PIC. Single photons are detected by four silicon-based SPADs (Excelitas), with the same efficiencies. Their output counts are collected by time-tagging electronics (Swabian Instruments) connected to a PC.
We used a total photon flux of the order of $10^6$ photons, with a time bin of $0.2$ $\mu$s set on the time tagger and acquired data for $30$ ms. For each measurement, ten separate acquisitions have been performed. Errors have then been estimated by considering the variance of such measurements. The data have been normalized with respect to $|100\rangle$ output chosen as a reference (efficiency $1$), thus giving the following output facet relative efficiencies: $0.487$ for $|000\rangle$, $0.975$ for $|010\rangle$, and $0.958$ for $|110\rangle$. The average dark counts of each SPAD have been removed from the raw data. No postselection operation is required to discard time bins where more than one photon in the setup has been detected. A Peltier Cell is used to thermally stabilize the PIC and is controlled by a PID controller. The circuit thermalizes for some seconds between each one of the ten acquisitions.

\backmatter



\bmhead{Acknowledgments}
We acknowledge helpful discussions with Claudio Conti, from University Sapienza, in the initial design of the experiment and with Jeongho Bang, from ETRI, during the preparation of the paper.


\bmhead{Funding}
This project has been supported by Q@TN, the joint lab between the University of Trento, FBK- Fondazione Bruno Kessler, INFN-National Institute for Nuclear Physics and CNR-National Research Council. This project has received funding from the European Union’s Horizon 2020 research and innovation programme under grant agreements No 820405 Project QRANGE and No 899368 Project EPIQUS.

\bmhead{Disclosures}
The authors declare no conflicts of interest.


\appendix

\section{The swap test algorithm}
\label{app:preliminaries}

\setcounter{equation}{0}
\renewcommand{\theequation}{A\arabic{equation}}
\setcounter{table}{0}
\renewcommand{\thetable}{A\arabic{table}}

The main purpose of the swap test \cite{swaptest} consists of evaluating the scalar product or the overlap between two unknown quantum states.
\begin{figure}[htbp]
\centering
\includegraphics{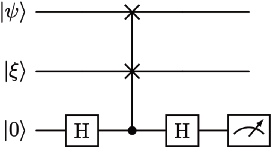}
\caption{\textbf{The swap test algorithm.} Gate representation of the swap test algorithm.}
\label{fig:swaptalgorithm}
\end{figure}
The swap test algorithm is represented in the quantum gate notation in Fig. \ref{fig:swaptalgorithm}(a). The circuit inputs are initialized to have the two qubits $| \psi \rangle $ and $| \xi \rangle$, whose scalar product we are interested in, and an ancillary qubit, also called ancilla, in the state $|0\rangle$. The first Hadamard gate (H) puts the ancillary qubit in a superposition of the states $|0\rangle$ and $|1\rangle$. Then, a controlled-swap gate (CSWAP) is applied to the total state with the auxiliary qubit as the control qubit: the action of this operation involves the swapping of the two target qubits when the auxiliary qubit is in the state $|1\rangle$. After this operation, the H gate is applied again to the auxiliary qubit. 
After some algebra, it can be shown that the circuit performs the following transformation after the second H gate:
\begin{equation}
    | \psi \rangle \otimes | \xi \rangle \otimes | 0 \rangle \to 
    \frac{1}{2} \left( | \psi \rangle \otimes | \xi \rangle + | \xi \rangle \otimes | \psi \rangle \right) \otimes | 0 \rangle + 
    \frac{1}{2} \left( | \psi \rangle \otimes | \xi \rangle - | \xi \rangle \otimes | \psi \rangle \right) \otimes | 1 \rangle \,.
\end{equation}
Finally, we measure the ancillary qubit state and the outcomes' probabilities $\{\mathbb{P}(x)\}_{x=0,1}$ are related to the scalar product of the two unknown states. In particular, the probabilities read:
\begin{equation}
    \mathbb{P}(x) = \frac{1}{2} \left( 1 + (-)^x |\langle \psi| \xi \rangle|^2 \right) 
    \,.
    \label{prob_result_swap}
\end{equation}
These relations can be easily inverted
\begin{equation}
    |\langle \psi| \xi \rangle|^2 = 2 \,\mathbb{P}(0) - 1 
    =
    1- 2 \,\mathbb{P}(1)
    \,.
    \label{prob_result_swap2}
\end{equation}
This result implies that the scalar product calculation is translated into a sampling problem: we need to acquire statistics about the outcomes of the auxiliary qubit in order to evaluate the overlap of the two qubits. 
The algorithm for the swap test described is usually called non-destructive because the measurement is done only on the ancilla and the remaining two qubits can be still used for other operations.

Consider now equation \eqref{prob_result_swap}. On the one hand, we can note that if the two qubit states $| \psi \rangle $ and $| \xi \rangle$ are equal or parallel, i.e. $|\langle \psi|\xi\rangle|^2 =1$, the auxiliary qubit has zero probability to be in the state $|1\rangle$. On the other hand, if the two qubits are orthogonal, i.e. $|\langle \psi|\xi\rangle|^2 = 0$, the auxiliary qubit has 50\% probability to be in the state $|0\rangle$ and $|1\rangle$. All the intermediate situations lie in the middle: as one can see from equations \eqref{prob_result_swap}-\eqref{prob_result_swap2}, in all the cases it holds that $0.5 \le \mathbb{P}(0) \le 1$ and $0 \le \mathbb{P}(1) \le 0.5$.
This algorithm can be implemented in the so-called 'quantum fingerprint' routine \cite{Buhrman_2001}: if two users, \textit{Alice} and \textit{Bob}, possess two secret keys, $x$ and $y$, and they want to compare them, they can send the keys to a reliable node, that through the swap test circuit can quantify how much the keys are similar. In this way, the equality problem, whose goal consists of finding
\begin{equation}
    f(x,y) =
    \begin{cases}
        1 & \mbox{if}\quad x = y \\
        0 & \mbox{if}\quad x \ne y 
    \end{cases}
\end{equation}
has an exponentially shorter encoding, given by the use of qubits \footnote{If both the users have $n$ bits, we have $O(\log n)$ qubits.}, and can be solved with a probability that depends on the number of the quantum hardware cycles.
The number of samples needed to obtain an error $\eta$ on the evaluation of the overlap with probability $\mathbb{P}(0) = 1-\delta$ scales as $O\!\left( ( 1-\delta)\delta\,\eta^{-2} \right)$ \cite{QSVM, scaling}. 

The swap test can be transformed in the SWITCH-test \cite{chamorroposada2023switch} to discriminate between two quantum evolutions $U_1$ and $U_2$ by setting the two unknown qubits $|\psi\rangle$ and $|\xi\rangle$ to $U_1|\chi\rangle$ and $U_2|\chi\rangle$, where $|\chi\rangle$ is a reference state. The result of the algorithm is
\begin{equation}
\mathbb{P}(x) = \frac{1}{2} \left( 1 + (-)^x |\langle \chi| U_1^\dagger U_2 | \chi \rangle|^2 \right) 
    \,,
\end{equation}
so the sampling can tell us how close the transformations $U_1$ and $U_2$ are.
If our goal is not given by just discrimination criteria but the detailed knowledge of a specific transformation, quantum Hamiltonian learning, proposed in \cite{Wiebe_2014,PhysRevA.89.042314} and implemented in \cite{Wang_2017}, it allows us to estimate the salient Hamiltonian parameters.

\section{Analytical description}
\label{app:theory}

To understand the state setting and manipulation in our swap test circuit, let's review some basic concepts about the manipulation of single photons in a set of waveguides. 
The fundamental building block for the preparation step is the MZI, which has 2 inputs and 2 outputs. The implemented MZI is formed by two sequential blocks, each one composed of a balanced multimode-interferometer-based integrated beam splitters (MMI) followed by phase shifters (PS) on each waveguide.
As usual, we can assign the vector $ 
\begin{pmatrix}
1 & 0
\end{pmatrix}^T$ 
to the top position or state $|0\rangle$ and the vector 
$\begin{pmatrix}
0 & 1
\end{pmatrix}^T$ to the down position or state $|1\rangle$.
Using this relation, the MMI and the PS are described by the following matrices \cite{Kok_2007}
\begin{equation}
U_{\rm MMI}= \frac{1}{\sqrt{2}}\begin{pmatrix}
1 & {\rm i} \\
{\rm i} & 1
\end{pmatrix}
\hspace{0.5cm},
\hspace{0.5cm}
U_{\rm PS}(\boldsymbol{\phi})= 
\begin{pmatrix}
{\rm e}^{{\rm i} \phi(1)} & 0 \\
0 & {\rm e}^{{\rm i} \phi(2)}
\end{pmatrix}
\label{BSandPS}
\end{equation}
and the MZI transformation reads
\begin{equation}
\begin{split}
U_{\rm MZI}(\boldsymbol{\theta})
&\equiv 
U_{\rm MMI} \cdot U_{\rm PS}(2\,\boldsymbol{\theta}) \cdot U_{\rm MMI} 
=
{\rm i} \,{\rm e}^{-{\rm i} (\theta(1)+\theta(2))}
\begin{pmatrix}
 \sin \Delta\theta 
 &\cos \Delta\theta \\
\cos \Delta\theta &
-\sin \Delta\theta 
\end{pmatrix} ,
\label{eqn:MZI_matrix}
\end{split}
\end{equation}
where $\Delta\theta=\theta(1)-\theta(2)$.
Starting from the initial state $|0\rangle$, and by applying the transformation
\begin{equation}
    U_{\rm rot}(\boldsymbol{\theta},\boldsymbol{\phi}) \equiv 
    U_{\rm PS}(\boldsymbol{\phi}) \cdot U_{\rm MZI}(\boldsymbol{\theta})
    \label{rot_matrix}
\end{equation}
we can reach any point on the Block sphere associated to the single qubit.
These transformations can then be embedded into the qudit structure, so the matrix representation of an MZI followed by PS that operates on the $k$-th and $(k+1)$-th waveguides reads \cite{Clements:16}
\begin{equation}
U_{\rm rot}^{(k)}(\boldsymbol{\theta},\boldsymbol{\phi}) \equiv
    \begin{pmatrix}
    1 & 0 &  \ldots & \ldots & \ldots & 0 \\
    0 & \ddots & & & &\vdots \\
    \vdots & & \left(U_{\rm {rot}}\right)_{11} & \left(U_{\rm {rot}}\right)_{12} & & \vdots \\
    \vdots & & \left(U_{\rm {rot}}\right)_{21} & \left(U_{\rm {rot}}\right)_{22} & & \vdots \\
    \vdots & & & & \ddots & 0 \\
    0 & \ldots & \ldots & \ldots & 0 & 1 \\
    \end{pmatrix} ,
\end{equation}
using equation \eqref{rot_matrix} for $U_{\rm {rot}}$ matrix entries and the following mapping to move, e.g. in the qu-quart case, from the Dirac notation to the vectorial notation: $
|00\rangle \iff \begin{pmatrix}
1 & 0 & 0 & 0
\end{pmatrix}^T$, 
$|01\rangle \iff \begin{pmatrix}
0 & 1 & 0 & 0
\end{pmatrix}^T$, 
$|10\rangle \iff \begin{pmatrix}
0 & 0 & 1 & 0
\end{pmatrix}^T$, 
$|11\rangle \iff \begin{pmatrix}
0 & 0 & 0 & 1
\end{pmatrix}^T$. 
For four waveguides we can have just three positions for a MZI followed by a PS, corresponding to $k=1,2,3$: these choices correspond to the cases where the MZI followed by a PS operates on the first and second, second and third, third and fourth waveguides, respectively.

Denoting the input phases of $U_{\rm {rot}}$s as a vector, the action of the triangular scheme of $U_{\rm {rot}}$s on the input photon states $|01\rangle$ is
\begin{equation}
\begin{split}
&U_{\rm rot}^{(2)}(\boldsymbol{\theta_{22}},\boldsymbol{\phi_{22}}) \cdot U_{\rm rot}^{(0)}(\boldsymbol{\theta_{21}},\boldsymbol{\phi_{21}}) \cdot U_{\rm rot}^{(1)}(\boldsymbol{\theta_1},\boldsymbol{\phi_1}) \cdot 
\begin{pmatrix}
0 \\ 1 \\ 0 \\ 0
\end{pmatrix}\\
&\hspace{1cm}=
-{\rm e}^{{\rm i} \sum_{j=1,2}(\theta_1(j)+\theta_{21}(j)+\theta_{22}(j))}
\begin{pmatrix}
{\rm e}^{{\rm i} (\phi_1(1) + \phi_{21}(1))} \sin \Delta\theta_1 \cos \Delta\theta_{21}\\
-{\rm e}^{{\rm i} (\phi_1(1) + \phi_{21}(2))} \sin \Delta\theta_1\sin \Delta\theta_{21}\\
{\rm e}^{{\rm i} ( \phi_1(2) + \phi_{22}(1))} \cos \Delta\theta_1 \sin \Delta\theta_{22}\\
{\rm e}^{{\rm i} ( \phi_1(2) + \phi_{22}(2))} \cos \Delta\theta_1 \cos \Delta\theta_{22}
\end{pmatrix} ,
\end{split}
\end{equation}
that can be written also in the Dirac notation as
\begin{equation}
\begin{split}
&U_{\rm rot}^{(2)}(\boldsymbol{\theta_{22}},\boldsymbol{\phi_{22}}) \cdot U_{\rm rot}^{(0)}(\boldsymbol{\theta_{21}},\boldsymbol{\phi_{21}}) \cdot U_{\rm rot}^{(1)}(\boldsymbol{\theta_1},\boldsymbol{\phi_1}) \cdot |01\rangle
\\
&\qquad= 
-{\rm e}^{{\rm i} \sum_{j=1,2}(\theta_1(j)+\theta_{21}(j)+\theta_{22}(j))}
\\
&\qquad\,\,\,\,\,
\left\{
{\rm e}^{{\rm i} (\phi_1(1) + \phi_{21}(1))} \sin \Delta\theta_{1} \left( \cos \Delta\theta_{21}|00\rangle
- {\rm e}^{-{\rm i} \Delta\phi_{21}} \sin \Delta\theta_{21} |01\rangle \right)
\right.\\
& \qquad\,\,\,\,\left.
+{\rm e}^{{\rm i} (\phi_1(2) + \phi_{22}(1))} \cos \Delta\theta_{1} \left( \sin \Delta\theta_{22}|10\rangle 
+ {\rm e}^{-{\rm i} \Delta\phi_{22}} \cos \Delta\theta_{22}|11\rangle \right) 
\right\}\, ,
\end{split}
\end{equation}
by defining $\Delta \phi_J \equiv \phi_J(1)- \phi_J(2)$ and using an abuse of notation in denoting the $U_{\rm rot}^{(k)}$ matrix and corresponding unitary transformation.
We can note from the previous two equations that the preparation step allows us to parameterize any state of two qubits, also entangled states: we can generically write any state of the form $ A_{00}|00\rangle + A_{01}|01\rangle + A_{10}|10\rangle + A_{11}|11\rangle $ with $|A_{00}|^2 + |A_{01}|^2 + |A_{10}|^2 + |A_{11}|^2 = 1$. 

The state is separable if $\boldsymbol{\theta_{2}}\equiv\boldsymbol{\theta_{21}}-\frac{\pi}{2}=\boldsymbol{\theta_{22}}$ and $\boldsymbol{\phi_{2}} \equiv \boldsymbol{\phi_{21}}=\boldsymbol{\phi_{22}}$, and, modulo a global phase, it reads
\begin{equation}
\begin{split}
&\left(\sin \Delta\theta_{1}|0\rangle + {\rm e}^{-{\rm i} \Delta\phi_{1}}\cos \Delta\theta_{1}|1\rangle\right)
\otimes
\left(\sin \Delta\theta_{2}|0\rangle + {\rm e}^{-{\rm i} \Delta\phi_{2}}\cos \Delta\theta_{2}|1\rangle \right)
\\
&\qquad\iff
\begin{pmatrix}
\sin\Delta\theta_{1} \sin \Delta\theta_{2}\\
{\rm e}^{-{\rm i} \Delta\phi_{2}}\sin \Delta\theta_{1} \cos \Delta\theta_{2}\\
{\rm e}^{-{\rm i} \Delta\phi_{1}}\cos \Delta\theta_{1} \sin \Delta\theta_{2}\\
{\rm e}^{-{\rm i} (\Delta\phi_{1}+\Delta\phi_{2})}\cos \Delta\theta_{1} \cos \Delta\theta_{2}
\end{pmatrix} ,
\label{eq:twoqubit}
\end{split}
\end{equation}
which is the generic form of a separable state of two qubits.
Therefore, the first MZI sets the first qubit, while the second and the third ones set the second qubit: the setting of the second qubit implies a defined relationship between the phases of the second and third MZIs \footnote{This point has important practical implications, since for a generic state it is required to compensate spurious global phase of the second and third MZIs due to fabrications imperfections.}.

Note that the squared scalar product of the two qubits present in the state written in equation \eqref{eq:twoqubit} reads
\begin{equation}
    \cos^2\left( \Delta\theta_{1}-\Delta\theta_{2}\right) \cos^2\left( \frac{\Delta\phi_{1}-\Delta\phi_{2}}{2}\right)
    +
    \cos^2\left( \Delta\theta_{1}+\Delta\theta_{2}
    \right) \sin^2\left( \frac{\Delta\phi_{1}-\Delta\phi_{2}}{2}\right)
    \,.
    \label{sca_pro_formula}
\end{equation}
We have shown that the preparation stage can be used to write down the two qubits whose overlap we want to evaluate and more generically the most general setting the three MZIs can span over all the possible configurations of the two-qubits Hilbert space.

At the end the preparation stage, other four waveguides are added. The nesting of the additional waveguides is done in agreement with the encoding of the two qubits written in the preparation stage. Moreover, the set with eight, or $2^3$, waveguides has the same structure of a tensor product of three qubits, where the first two qubits are the inputs of the swap test and the third is the ancilla. Since only the even waveguides are connected to the previous qu-quart structure, in our circuit the ancillary qubit is always set to the state $|0\rangle$ by design. 
The map between the qu-eighth states and the three-qubit states reads as 
\begin{align}
    &|0\rangle \to |000\rangle \equiv |0\rangle\otimes|0\rangle\otimes|0\rangle\,, \qquad
     |1\rangle \to |001\rangle \equiv |0\rangle\otimes|0\rangle\otimes|1\rangle\,, 
     \nonumber\\
     & |2\rangle \to |010\rangle \equiv |0\rangle\otimes|1\rangle\otimes|0\rangle\,, \qquad
     \mbox{etc.} \,,
\end{align}
where the first two positions belong to the qubits, whose scalar product we want to calculate, and the third is the ancillary qubit.

After the preparation stage, we have the swap test stage. This step starts with a set of MMIs between every even and odd waveguide: this is equivalent to applying the transformation $\mathds{1} \otimes \mathds{1} \otimes U_{\rm MMI} $ to the state $|\psi\rangle \otimes |\xi\rangle \otimes |0\rangle$. We are operating just on the ancillary qubit as it is described in Supplementary Information Section \ref{app:preliminaries}: in our case, instead of using an H gate we use the transformation $U_{\rm MMI}$. The action of this first step, given by the transformation $\mathds{1} \otimes \mathds{1} \otimes U_{\rm MMI} $, can be written in matrix notation as
\begin{align}
\frac{1}{\sqrt{2}}
\begin{pmatrix}
1 & {\rm i} & 0 & 0 & 0 & 0 & 0 & 0 \\
{\rm i} & 1 & 0 & 0 & 0 & 0 & 0 & 0 \\
0 & 0 & 1 & {\rm i} & 0 & 0 & 0 & 0 \\
0 & 0 & {\rm i} & 1 & 0 & 0 & 0 & 0 \\
0 & 0 & 0 & 0 & 1 & {\rm i} & 0 & 0 \\
0 & 0 & 0 & 0 & {\rm i} & 1 & 0 & 0 \\
0 & 0 & 0 & 0 & 0 & 0 & 1 & {\rm i} \\
0 & 0 & 0 & 0 & 0 & 0 & {\rm i} & 1 
\end{pmatrix} ,
\label{BSonall_matrix}
\end{align}
where we using again the following map to move from the Dirac notation to the vectorial notation:
$
|000\rangle \iff \begin{pmatrix}
1 & 0 & 0 & 0 & 0 & 0 & 0 & 0
\end{pmatrix}^T$, 
$|001\rangle \iff \begin{pmatrix}
0 & 1 & 0 & 0 & 0 & 0 & 0 & 0
\end{pmatrix}^T$, $|010\rangle \iff \begin{pmatrix}
0 & 0 & 1 & 0 & 0 & 0 & 0 & 0
\end{pmatrix}^T$, etc.
After this step, we have a central network of waveguide crossings to implement $U_{\rm CSWAP}$, which has the following matrix representation
\begin{align}
\begin{pmatrix}
1 & 0 & 0 & 0 & 0 & 0 & 0 & 0 \\
0 & 1 & 0 & 0 & 0 & 0 & 0 & 0 \\
0 & 0 & 1 & 0 & 0 & 0 & 0 & 0 \\
0 & 0 & 0 & 0 & 0 & 1 & 0 & 0 \\
0 & 0 & 0 & 0 & 1 & 0 & 0 & 0 \\
0 & 0 & 0 & 1 & 0 & 0 & 0 & 0 \\
0 & 0 & 0 & 0 & 0 & 0 & 1 & 0 \\
0 & 0 & 0 & 0 & 0 & 0 & 0 & 1
\end{pmatrix} .
\label{cswap_matrix}
\end{align}
Looking at the eight basis vectors, it is easy to recognize that the only non-trivial action of the CSWAP consists of the swapping of the states $|011\rangle$ and $|101\rangle$, whose action is implemented by the non-diagonal terms of the previous matrix. 
Finally, we insert phase shifters on all eight waveguides to compensate for spurious phases and another layer of MMI to implement the final transformation $\mathds{1} \otimes \mathds{1} \otimes U_{\rm MMI} $ before the detection. 
The matrix for the step with the phase shifters can be written as a diagonal matrix with different phases in the entries: we denote it by $U_{\text{PS}_8}$ and represent it as the matrix $ \{ {\rm e}^{{\rm i}\theta_s(k)} \delta_{k h} \}_{k,h=1..8}$.

We point out the fact that our implementation of the swap test algorithm on the PIC is in one-to-one correspondence with the circuit reported in Fig. \ref{fig:swaptalgorithm}, provided we change the H gate with $U_{\rm MMI}$, equation \eqref{BSandPS}. From this consideration, we can already understand that in our case the $\mathbb{P}(0)$ and $\mathbb{P}(1)$ are switched with respect to the standard swap test result given in equation \eqref{prob_result_swap}: this is due to the fact that ${\rm H}^2 = \mathds{1}$ and $U_{\rm MMI}^2 = {\rm i}\,\sigma_x$, where $\sigma_x$ is the Pauli matrix $\footnotesize \begin{pmatrix}
0 & 1 \\
1 & 0
\end{pmatrix}$.

The manipulation stage for the swap test can be represented as
\begin{align}
U_{\rm swaptest}=\left(\mathds{1} \otimes \mathds{1} \otimes U_{\rm MMI}\right) \cdot 
U_{\text{PS}_8} (\boldsymbol{\theta_s})
\cdot U_{\rm CSWAP} \cdot
\left(\mathds{1} \otimes \mathds{1} \otimes U_{\rm MMI}\right) \,.
\end{align}
By using Eqs. \eqref{BSonall_matrix} - \eqref{cswap_matrix} and the phases matrix $U_{\text{PS}_8}$, the matrix describing the manipulation stage can be factorized in three mutually commuting contributions as
\begin{align}
\textbf{MZI}^{(0)}\left(\theta_s(1),\theta_s(2)\right) \cdot
\textbf{MZI}^{(6)}\left(\theta_s(7),\theta_s(8)\right) \cdot
U_\textbf{swap-core}\left( \theta_s(3),\theta_s(4),\theta_s(5),\theta_s(6) \right) \,,
\label{eq:swap_part}
\end{align}
where $U_\textbf{swap-core}$ is defined as
\begin{align}
U_\textbf{swap-core}\!\left( \theta_s(3),\theta_s(4),\theta_s(5),\theta_s(6) \right) \equiv
\frac{1}{2}\!\!
\begin{pmatrix}
2 & 0 & 0 & 0 & 0 & 0 & 0 & 0 \\
0 & 2 & 0 & 0 & 0 & 0 & 0 & 0 \\
0 & 0 & {\rm e}^{{\rm i} \theta_s(3)} & {\rm i}\, {\rm e}^{{\rm i} \theta_s(3)} & -{\rm e}^{{\rm i} \theta_s(4)} & {\rm i}\, {\rm e}^{{\rm i} \theta_s(4)} & 0 & 0 \\
0 & 0 & {\rm i}\, {\rm e}^{{\rm i} \theta_s(3)} & -{\rm e}^{{\rm i} \theta_s(3)} & {\rm i}\, {\rm e}^{{\rm i} \theta_s(4)} & {\rm e}^{{\rm i} \theta_s(4)} & 0 & 0 \\
0 & 0 &  -{\rm e}^{{\rm i} \theta_s(6)} & {\rm i}\, {\rm e}^{{\rm i} \theta_s(6)} & {\rm e}^{{\rm i} \theta_s(5)} & {\rm i}\, {\rm e}^{{\rm i} \theta_s(5)} & 0 & 0 \\
0 & 0 & {\rm i}\, {\rm e}^{{\rm i} \theta_s(6)} & {\rm e}^{{\rm i} \theta_s(6)} & {\rm i}\, {\rm e}^{{\rm i} \theta_s(5)} & -{\rm e}^{{\rm i} \theta_s(5)} & 0 & 0\\
0 & 0 & 0 & 0 & 0 & 0 & 2 & 0 \\
0 & 0 & 0 & 0 & 0 & 0 & 0 & 2
\end{pmatrix} \!.
\label{swap-core}
\end{align}
The three contributions represent the MZI between the first and second waveguide, the central part from the third waveguide to the sixth waveguide and the MZI between the seventh and eighth waveguide, respectively. Interestingly, the 4x4 central part of $U_\textbf{swap-core}$ has the same matrix of a 4x4 balanced MMI whose internal phases can be tuned \cite{Peruzzo_2011}. 

Setting all the phases $\boldsymbol{\theta_s}$ to zero and applying it to the generic state set in the preparation stage given in equation \eqref{eq:twoqubit} embedded in the qu-eighth structure, we obtain
\begin{align}
&\begin{pmatrix}
0 & {\rm i} & 0 & 0 & 0 & 0 & 0 & 0 \\
{\rm i} & 0 & 0 & 0 & 0 & 0 & 0 & 0 \\
0 & 0 & \frac{1}{2} & \frac{{\rm i}}{2} & -\frac{1}{2} & \frac{{\rm i}}{2} & 0 & 0 \\
0 & 0 & \frac{{\rm i}}{2} & -\frac{1}{2} & \frac{{\rm i}}{2} & \frac{1}{2} & 0 & 0 \\
0 & 0 &  -\frac{1}{2} & \frac{{\rm i}}{2} & \frac{1}{2} & \frac{{\rm i}}{2} & 0 & 0 \\
0 & 0 & \frac{{\rm i}}{2} & \frac{1}{2} & \frac{{\rm i}}{2} & -\frac{1}{2} & 0 & 0\\
0 & 0 & 0 & 0 & 0 & 0 & 0 & {\rm i} \\
0 & 0 & 0 & 0 & 0 & 0 & {\rm i} & 0
\end{pmatrix}
\cdot
\begin{pmatrix}
\sin\Delta\theta_{1} \sin \Delta\theta_{2}\\
0\\
{\rm e}^{-{\rm i} \Delta\phi_{2}}\sin \Delta\theta_{1} \cos \Delta\theta_{2}\\
0\\
{\rm e}^{-{\rm i} \Delta\phi_{1}}\cos \Delta\theta_{1} \sin \Delta\theta_{2}\\
0\\
{\rm e}^{-{\rm i} (\Delta\phi_{1}+\Delta\phi_{2})}\cos \Delta\theta_{1} \cos \Delta\theta_{2}\\
0
\end{pmatrix}
\nonumber\\
& \hspace{3cm}=\frac{1}{2}
\begin{pmatrix}
0\\
2{\rm i}\,\sin\Delta\theta_{1} \sin \Delta\theta_{2}\\
{\rm e}^{-{\rm i} \Delta\phi_{2}}\sin \Delta\theta_{1} \cos \Delta\theta_{2}
- {\rm e}^{-{\rm i} \Delta\phi_{1}}\cos \Delta\theta_{1} \sin \Delta\theta_{2}\\
{\rm i}\,{\rm e}^{-{\rm i} \Delta\phi_{2}}\sin \Delta\theta_{1} \cos \Delta\theta_{2}
+ {\rm i}\,{\rm e}^{-{\rm i} \Delta\phi_{1}}\cos \Delta\theta_{1} \sin \Delta\theta_{2}\\
-{\rm e}^{-{\rm i} \Delta\phi_{2}}\sin \Delta\theta_{1} \cos \Delta\theta_{2}
+{\rm e}^{-{\rm i} \Delta\phi_{1}}\cos \Delta\theta_{1} \sin \Delta\theta_{2}\\
{\rm i}\,{\rm e}^{-{\rm i} \Delta\phi_{2}}\sin \Delta\theta_{1} \cos \Delta\theta_{2}
+{\rm i}\,{\rm e}^{-{\rm i} \Delta\phi_{1}}\cos \Delta\theta_{1} \sin \Delta\theta_{2}\\
0\\
2{\rm i}\,{\rm e}^{-{\rm i} (\Delta\phi_{1}+\Delta\phi_{2})}\cos \Delta\theta_{1} \cos \Delta\theta_{2}
\end{pmatrix} 
.
\end{align}
The values for $\{\mathbb{P}(x)\}_{x=0,1}$ are given by the sum of the squared modulus of even entries of the previous vector for $\mathbb{P}(0)$ and the sum of the squared modulus of odd entries of the previous vector for $\mathbb{P}(1)$.
In particular, the values for $\{\mathbb{P}(x)\}_{x=0,1}$ are
\begin{equation}
\begin{split}
\mathbb{P}(0) = \frac{1}{2} \!\left( \!1 \!-\! 
\cos^2\!\left( \Delta\theta_{1} \!-\! \Delta\theta_{2}\right) 
\cos^2\!\left( \!\frac{\Delta\phi_{1} \!-\! \Delta\phi_{2}}{2}\!\right)
\!-\!
\cos^2\!\left( \Delta\theta_{1} \!+\! \Delta\theta_{2} \right) 
\sin^2\!\left( \!\frac{\Delta\phi_{1} \!-\! \Delta\phi_{2}}{2}\!\right) \!\right) \!,
\\
\mathbb{P}(1) = \frac{1}{2} \!\left( \!1 \!+\! 
\cos^2\!\left( \Delta\theta_{1} \!-\! \Delta\theta_{2}\right) 
\cos^2\!\left( \!\frac{\Delta\phi_{1} \!-\! \Delta\phi_{2}}{2}\!\right)
\!+\!
\cos^2\!\left( \Delta\theta_{1} \!+\! \Delta\theta_{2}\right) 
\sin^2\!\left( \!\frac{\Delta\phi_{1} \!-\! \Delta\phi_{2}}{2}\!\right) \!\right) \!.
\end{split}
\end{equation}


We conclude this section by showing how it is possible to swap the four outputs corresponding to the state $|0\rangle$ of the ancilla with the outputs corresponding to the $|1\rangle$ of the ancilla using the phase shifters present in the swap test scheme.
In this way the sampling procedure can be achieved with four detectors with two runs.
Swapping $|000\rangle \iff |001\rangle$ and $|110\rangle \iff |111\rangle$ is very easy, since they are connected by a MZI, equation \eqref{eq:swap_part}. Thus, we can simply tune two phases to achieve this.
Finally, let's analyse the central part of the swap test circuit composed by waveguides corresponding to states $|010\rangle, |011\rangle, |100\rangle, |101\rangle$. 
This part has four inputs and four outputs and it is described by the following matrix

\begin{align}
& \frac{1}{2}
\begin{pmatrix}
1&{\rm i}&0&0\\
{\rm i}&1&0&0\\
0&0&1&{\rm i}\\
0&0&{\rm i}&1
\end{pmatrix}
\begin{pmatrix}
{\rm e}^{{\rm i} \theta_s(3)}&0&0&0\\
0&{\rm e}^{{\rm i} \theta_s(4)}&0&0\\
0&0&{\rm e}^{{\rm i} \theta_s(5)}&0\\
0&0&0&{\rm e}^{{\rm i} \theta_s(6)}
\end{pmatrix}
\begin{pmatrix}
1&0&0&0\\
0&0&0&1\\
0&0&1&0\\
0&1&0&0
\end{pmatrix}
\begin{pmatrix}
1&{\rm i}&0&0\\
{\rm i}&1&0&0\\
0&0&1&{\rm i}\\
0&0&{\rm i}&1
\end{pmatrix}
\nonumber\\
& \qquad=\frac{1}{2}
\begin{pmatrix}
{\rm e}^{{\rm i} \theta_s(3)} & {\rm i}\, {\rm e}^{{\rm i} \theta_s(3)} & -{\rm e}^{{\rm i} \theta_s(4)} & {\rm i}\, {\rm e}^{{\rm i} \theta_s(4)} \\
{\rm i}\, {\rm e}^{{\rm i} \theta_s(3)} & -{\rm e}^{{\rm i} \theta_s(3)} & {\rm i}\, {\rm e}^{{\rm i} \theta_s(4)} & {\rm e}^{{\rm i} \theta_s(4)} \\
-{\rm e}^{{\rm i} \theta_s(6)} & {\rm i}\, {\rm e}^{{\rm i} \theta_s(6)} & {\rm e}^{{\rm i} \theta_s(5)} & {\rm i}\, {\rm e}^{{\rm i} \theta_s(5)} \\
{\rm i}\, {\rm e}^{{\rm i} \theta_s(6)} & {\rm e}^{{\rm i} \theta_s(6)} & {\rm i}\, {\rm e}^{{\rm i} \theta_s(5)} & -{\rm e}^{{\rm i} \theta_s(5)}
\end{pmatrix} ,
\label{eq:centralSWAP}
\end{align}
which is the central submatrix of $U_\textbf{swap-core}$ reported in equation \eqref{swap-core}.
In such a central structure the generic state that we can inject is $(A,0,B,0)$, since the ancilla is always set to the state $|0\rangle$. The result is
\begin{align}
\frac{1}{2}
\begin{pmatrix}
A {\rm e}^{{\rm i}\theta_s(3)}-B {\rm e}^{{\rm i}\theta_s(4)}\\
{\rm i}\left(A {\rm e}^{{\rm i}\theta_s(3)}+B {\rm e}^{{\rm i}\theta_s(4)}\right)\\
-A {\rm e}^{{\rm i}\theta_s(6)}+B {\rm e}^{{\rm i}\theta_s(5)}\\
{\rm i}\left(A {\rm e}^{{\rm i}\theta_s(6)}+B {\rm e}^{{\rm i}\theta_s(5)}\right)
\end{pmatrix} .
\end{align}
Inverting the computational-basis states $|0\rangle \iff |1\rangle$ for the ancilla is equivalent to the swapping of the first with the second entry and of the third with the fourth entry
\begin{align}
\frac{1}{2}
\begin{pmatrix}
{\rm i}\left(A {\rm e}^{{\rm i}\theta_s(3)}+B {\rm e}^{{\rm i}\theta_s(4)}\right)\\
A {\rm e}^{{\rm i}\theta_s(3)}-B {\rm e}^{{\rm i}\theta_s(4)}\\
{\rm i}\,\left(A {\rm e}^{{\rm i}\theta_s(6)}+B {\rm e}^{{\rm i}\theta_s(5)}\right)\\
-A {\rm e}^{{\rm i}\theta_s(6)}+B {\rm e}^{{\rm i}\theta_s(5)}
\end{pmatrix} .
\label{eq:cenpart1}
\end{align}
Sending $(\theta_s(3), \theta_s(4), \theta_s(5), \theta_s(6)) \to (\theta_s(3), \theta_s(4)+\pi, \theta_s(5), \theta_s(6)+\pi)$, we have
\begin{align}
\frac{1}{2}
\begin{pmatrix}
A {\rm e}^{{\rm i}\theta_s(3)}+B {\rm e}^{{\rm i}\theta_s(4)}\\
{\rm i}\left(A {\rm e}^{{\rm i}\theta_s(3)}-B {\rm e}^{{\rm i}\theta_s(4)}\right)\\
A {\rm e}^{{\rm i}\theta_s(6)}+B {\rm e}^{{\rm i}\theta_s(5)}\\
{\rm i}\left(-A {\rm e}^{{\rm i}\theta_s(6)}+B {\rm e}^{{\rm i}\theta_s(5)}\right)
\end{pmatrix} .
\label{eq:cenpart2}
\end{align}
We can note that each entry modulus of \eqref{eq:cenpart1} is equal to each entry modulus of \eqref{eq:cenpart2}: since after this operation we measure, we can ignore the phase difference and affirm that swapping the computational-basis states of the ancilla is equivalent to sending $(\theta_s(3), \theta_s(4), \theta_s(5), \theta_s(6)) \to (\theta_s(3), \theta_s(4)+\pi, \theta_s(5), \theta_s(6)+\pi)$. Therefore, we can answer positively to the question, 'can we put detectors on just half of the exits?'.

\section{Model for PIC's real characteristics}
\label{app:ni_model}
A model taking into account the real characteristics of our PIC is necessary for a better understanding of our swap test integrated photonic hardware, whose outputs can considerably deviate from what expected theoretically. A similar analysis has already been carried out in \cite{Mazzucchi2021,Leone2022,Leone2023}, where further information about our procedure can be found. Here in particular, we consider that:
\begin{itemize}
    \item[1] MMIs are not exactly implementing the matrix of an ideal beam splitter due to fabrication errors;
    \item[2] CRs do not have a perfectly unitary transfer function;
    \item[3] calibration errors and power supply errors affect the exact values of the currents to which every PS is nominally set.
\end{itemize}
To consider point 1, the matrix representation of an MMI becomes:
\begin{equation}
U_{\text{MMI}}(t,r)=
\begin{pmatrix}
 t& {\rm i} r \\
 {\rm i} r  & t\\
\end{pmatrix},
\end{equation}
where $t \neq r \neq \frac{1}{\sqrt{2}}$. Note that in principle, $t^2+r^2\leq 1$. To explicitly consider it, we rewrite the operator as:
\begin{align}
&U_{\text{MMI}}(\alpha,\beta)=\beta
\begin{pmatrix}
 \cos(\alpha)& {\rm i} \sin(\alpha) \\
 {\rm i} \sin(\alpha)  & \cos(\alpha)\\
\end{pmatrix},\\
&\beta=\sqrt{t^2+r^2}\,\,, \qquad \cos(\alpha)=\frac{t}{\sqrt{t^2+r^2}}\,\,, \qquad \sin(\alpha)=\frac{r}{\sqrt{t^2+r^2}}\,\,.
\end{align}  

For point 2, we model our crossing region in the swap test as:
 \begin{equation}
U_{\text{CSWAP}}(T)=
\sqrt{T}\begin{pmatrix} 1&  0&   0&   0&   0&   0&   0&   0\\
        0&   1& 0&   0&   0&   0&   0&   0\\
        0&   0&   1& 0&   0&   0&   0&   0\\
        0&   0&   0&   0&   0&   1&   0&   0\\
        0&   0&   0&   0&   1&   0&   0&   0\\
        0&   0&   0&   1&   0&   0&   0&   0\\
        0&   0&   0&   0&   0&   0&   1& 0\\
        0&   0&   0&   0&   0&   0&   0&   1
        \end{pmatrix},
\end{equation}
where $T<1$ is the power transmission of a single crossing component.

To consider point 3, we add error terms in every matrix representation of the PS. We introduce the notation $U_{\text{PS}_n}(\boldsymbol{\theta},\boldsymbol{\delta})$, defined as:
\begin{equation}
U_{\text{PS}_n}(\boldsymbol{\theta},\boldsymbol{\delta})=
\begin{pmatrix} 
        {\rm e}^{ {\rm i}( \theta(1)+\delta(1))}&  \ldots  &   0\\
        \vdots&   \ddots&  \vdots   \\
        0&  \ldots &   {\rm e}^{ {\rm i}( \theta(n)+\delta(n))}
        \end{pmatrix}.\\
\end{equation}
Note that the terms $\boldsymbol{\delta}$ only consider the power supply instabilities and the error introduced by the fitting operation for the calibration on the different MZIs and PSs of the PIC. The thermal crosstalk, which in principle acts in the same way by introducing further phase error terms, here it is not considered because it is quite difficult to quantify its effect on the different PSs.

Before introducing the action of the different stages of the PIC, we introduce the action of a single MZI, which modifies equation \eqref{eqn:MZI_matrix} as:
\begin{equation}            
U_{\text{MZI}}(\boldsymbol{\theta},\boldsymbol{\delta})=U_{\text{MMI}}(\alpha,\beta) \cdot U_{\text{PS}_2}(2\boldsymbol{\theta},2\boldsymbol{\delta})\cdot U_{\text{MMI}}\,.
\end{equation}
Note that in the previous equation, we have dropped the explicit dependence on $(\alpha,\beta)$ for clarity reasons.
The matrix connected to the projection operation to the state $|x\rangle_{x=0,1}$ is instead given by:
\begin{equation}
P_x=
\begin{pmatrix}
 1-x & 0 \\
 0 & x \\
\end{pmatrix}.\\\
\end{equation}
The action of the preparation stage is described by:
\begin{equation}
    \begin{aligned}
&U_{\text{preparation}}(\boldsymbol{\theta_{\text{Q1}}},\boldsymbol{\theta_{\text{Q2}}},\boldsymbol{\phi_{\text{Q1}}},\boldsymbol{\phi_{\text{Q2}}},\boldsymbol{\delta_{\text{Q1}}},\boldsymbol{\delta_{\text{Q2}_1}},\boldsymbol{\delta_{\text{Q2}_2}},\boldsymbol{\delta_{\text{Q2}_3}})\\
&\qquad =U_{\text{setQ2}}(\boldsymbol{\theta_{\text{Q2}}},\boldsymbol{\phi_{\text{Q2}}},\boldsymbol{\delta_{\text{Q2}_1}},\boldsymbol{\delta_{\text{Q2}_2}},\boldsymbol{\delta_{\text{Q2}_3}})
\cdot U_{\text{setQ1}}(\boldsymbol{\theta_{\text{Q1}}},\boldsymbol{\phi_{\text{Q1}}},\boldsymbol{\delta_{\text{Q1}}})\,.
 \end{aligned}
\end{equation}
The operator $U_{\text{setQ1}}$ and $U_{\text{setQ2}}$, respectively, set the first and the second qubit of our state and are described as:
 \begin{equation}
    \begin{aligned}  
    &U_{\text{setQ1}}(\boldsymbol{\theta_{\text{Q1}}},\boldsymbol{\phi_{\text{Q1}}},\boldsymbol{\delta_{\text{Q1}}})=\\
   &\left(P_0 \otimes \sigma_x\otimes\mathds{1}+P_1\otimes\mathds{1}\otimes\mathds{1}\right) \cdot (U_{\text{PS}_2}(\boldsymbol{\phi_{\text{Q1}}},\boldsymbol{\delta_{\text{Q1}}})\otimes\mathds{1}\otimes\mathds{1})\cdot
            \left(U_{\text{MZI}}(\boldsymbol{\theta_{\text{Q1}}},\boldsymbol{\delta_{\text{Q1}}})\otimes\mathds{1}\otimes\mathds{1} \right)\,,\\
\end{aligned}
\end{equation}  
and
 \begin{equation}
    \begin{aligned}
    &U_{\text{setQ2}}(\boldsymbol{\theta_{\text{Q2}}},\boldsymbol{\phi_{\text{Q2}}},\boldsymbol{\delta_{\text{Q2}_1}},\boldsymbol{\delta_{\text{Q2}_2}},\boldsymbol{\delta_{\text{Q2}_3}})=\\
    &\!(U_{\text{PS}_4}(\boldsymbol{\phi_{\text{Q2}}},\boldsymbol{\delta_{\text{Q2}_3}})\otimes\mathds{1})
    \!\cdot\!
    \left( \! P_0\otimes U_{\text{MZI}}(\boldsymbol{\theta_{\text{Q2}}}-\frac{\pi}{2},\boldsymbol{\delta_{\text{Q2}_2}})\otimes \mathds{1})\!+\!P_1\otimes U_{\text{MZI}}(\boldsymbol{\theta_{\text{Q2}}},\boldsymbol{\delta_{\text{Q2}_1}})\otimes\mathds{1}))\!\right)\!,\\
\end{aligned}
\end{equation}
The action of the swap test can be described as:
\begin{equation}
    \begin{aligned}
        &U_{\text{swaptest}}(\boldsymbol{\phi_s},\boldsymbol{\delta_s})= \left(\mathds{1} \otimes \mathds{1} \otimes U_{\text{MMI}} \right) \cdot U_{\text{PS}_8} (\boldsymbol{\phi_s},\boldsymbol{\delta_s})\cdot U_{\text{CSWAP}} \cdot \left( \mathds{1} \otimes \mathds{1} \otimes U_{\text{MMI}} \right),
    \end{aligned}
\end{equation}
Eventually, the whole action of the PIC can be written as:
\begin{equation}
\begin{aligned}   &U(\boldsymbol{\theta_{\text{Q1}}},\boldsymbol{\theta_{\text{Q2}}},\boldsymbol{\phi_{\text{Q1}}},\boldsymbol{\phi_{\text{Q2}}},\boldsymbol{\phi_s},\boldsymbol{\delta_{\text{Q1}}},\boldsymbol{\delta_{\text{Q2}_1}},\boldsymbol{\delta_{\text{Q2}_2}},\boldsymbol{\delta_{\text{Q2}_3}},\boldsymbol{\delta_s})=\\
&U_{\text{swaptest}}(\boldsymbol{\phi_s},\boldsymbol{\delta_s})
\cdot U_{\text{preparation}}(\boldsymbol{\theta_{\text{Q1}}},\boldsymbol{\theta_{\text{Q2}}},\boldsymbol{\phi_{\text{Q1}}},\boldsymbol{\phi_{\text{Q2}}},\boldsymbol{\delta_{\text{Q1}}},\boldsymbol{\delta_{\text{Q2}_1}},\boldsymbol{\delta_{\text{Q2}_2}},\boldsymbol{\delta_{\text{Q2}_3}}) \,.\\
\end{aligned}
\end{equation}
    
Note that in the previous equations, we have omitted the dependence of the different operators from $\alpha$ and $T$ for clarity. 
Consequently, the output density matrix at the end of our PIC is given by:
 \begin{equation}
     \rho_{\text{out}}=U\rho_{\text{in}}U^\dag \,,
 \end{equation}
 where $\rho_{in}=|010\rangle\langle010|$.
The probability of observing the outcome $x$ in the third qubit at the output of the PIC can then be written as:
\begin{equation}
\label{app:eq_prob}
    \mathbb{P}(x)=\frac{ \Tr[U\rho_{\text{in}} U^\dag \cdot \left(\mathds{1} \otimes \mathds{1} \otimes P_x \right) ]}{\Tr[U\rho_{\text{in}} U^\dag ] } \, .
\end{equation}
The normalization term $\Tr[U\rho_{\text{in}} U^\dag ]$ is necessary due to the presence of the non-unitary operator $U_{\text{MMI}}(\alpha,\beta)$ and $U_{\text{CSWAP}}(T)$. Indeed, this can be handled as done in \cite{Leone2022} since we are using a linear system, which acts directly at the single-photon level. Moreover, due to this normalization, one can set the values of $\beta$ and $T$ to 1 as, being just multiplicative factors appearing in both numerator and denominator of equation \eqref{app:eq_prob}, they cancel out.
Once obtained the probabilities $\{\mathbb{P}(x)\}_{x=0,1}$, then the scalar product is calculated according to equation \eqref{app:eq_prob}.
The $n\sigma$ confidence intervals on the scalar product is calculated by applying the error propagation code reported in \cite{sofwareUncertainty} considering $n\sigma$ errors on the measured quantities ($\alpha$, $1\sigma_\alpha=0.02$, $T$, $1\sigma_T=0.002$) and on the estimated ones (all the phases terms). Note that we have modified the solver to the Matlab pattern-search algorithm with respect to the code reported in \cite{sofwareUncertainty}.

\section{Use of an attenuated source}
\label{app:att_source}
The original discussion of this situation can be found in \cite{Pasini2020}.
In the case of an attenuated laser source, the input state at the input of the PIC is 
\begin{equation}\label{state-c}
    |\Psi \rangle = \sum_{n=0}^{+\infty} C_n |n_{|010\rangle} \rangle\:,\quad \mbox{where} \quad \sum_{n=0}^{+\infty} |C_n|^2 =1 \,.
\end{equation}
$|n_{|010\rangle} \rangle = \frac{1}{\sqrt{n!}} (a_{|010\rangle}^\dagger)^n |\text{vac}\rangle$ is the number state which have $n$ single photons in the state $|010\rangle$, while $|\text{vac}\rangle$ is the vacuum state and $a_{|010\rangle}^\dagger$ is the creation operator of a photon in the state $|\psi\rangle$. \\
Now, the action of the preparation state is to prepare the state $|\psi\xi0\rangle$, starting from the initial state $|010\rangle$:
\begin{equation}
	|010\rangle  \to  U_{\text{preparation}}|010\rangle = |\psi\xi0\rangle \,.
\end{equation}
Similarly, the action of the swap test can be described in the same way:
\begin{equation}
	|\psi\xi0\rangle \to U_{\text{swaptest}}|\psi\xi0\rangle=|\zeta\rangle \,.
\end{equation}
Taking into account that in our PIC every element is linear, the cumulative action of the PIC can be described by the unitary operator $U=
U_{\text{swaptest}}
\cdot U_{\text{preparation}}$ for which the following requirement must be respected:
\begin{itemize}
	\item the number of photons is conserved, i.e. $U a_{|010\rangle}^\dagger (U)^\dagger = a^\dagger_{\zeta}$,
	\item the vacuum is invariant under the operator $U$, i.e. $U|\text{vac}\rangle = |\text{vac}\rangle$.
\end{itemize}
Therefore, the net action of the PIC on the input state $|\Psi\rangle$ is:
\begin{equation} 
		|\Psi \rangle \to |\Psi^\prime \rangle = U |\Psi \rangle = \sum_{n=0}^{+\infty} C_n |n_{\zeta} \rangle \,.
\end{equation}
Considering the SPADs detectors, it is possible to apply the same discussion reported in \cite{Pasini2020}. As long as our detectors can fairly sample the probability distribution of the photons whose  arrival time is not important, the detection events associated with diﬀerent photons of the same pulse can be treated as independent identically distributed discrete random variables with probabilities given by equation \eqref{app:eq_prob}. Consequently, the collected data allows us to estimate the square scalar products correctly. Note that this discussion also holds for mixed input state $\rho_{\rm in}$ instead of $|\Psi\rangle$.

\section{Components characterization}
\label{app:ChAR}
\begin{figure}
\centering
  \includegraphics[width=\textwidth]{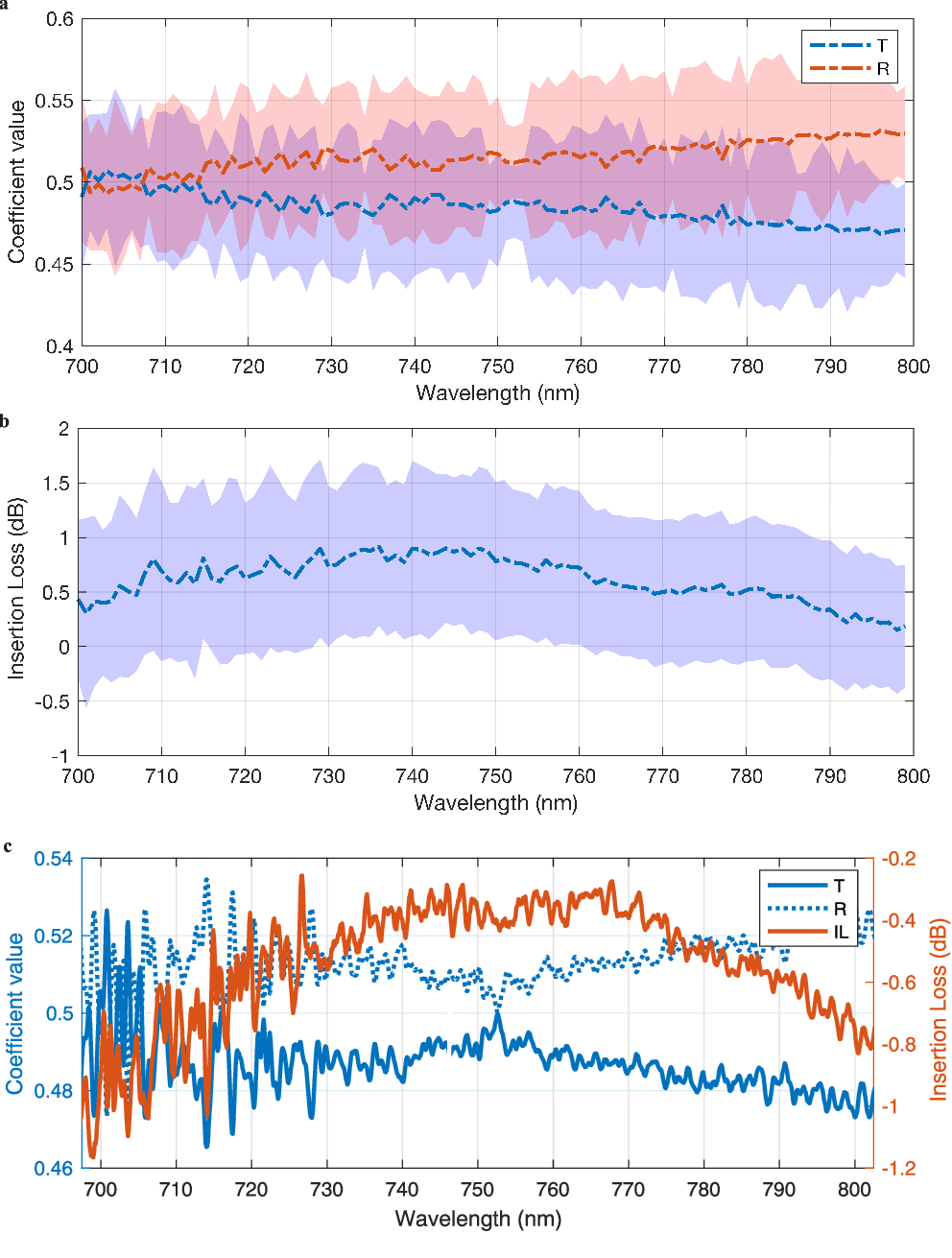}
\caption{\textbf{Characterization curves of MMI components.} 
\textbf{a)} Example of the characterization curves for a single MMI. In blue, it is reported the mean normalized transmission coefficient T, while in red, it is reported the mean normalized reflection coefficient R. The colored areas represent the $2\sigma$ confidence interval for the two coefficients. \textbf{b)} Mean value of the insertion loss measured for the different chips. The colored area represents the $2\sigma$ confidence interval. \textbf{c)} Transmission (solid blue, T), reflection (dotted blue, R) and insertion loss (red, IL) coefficients for the MMI in one of the chips.}
\label{fig:charmmi_app}
\end{figure}

\begin{figure}
\centering
  \includegraphics[width=\textwidth]{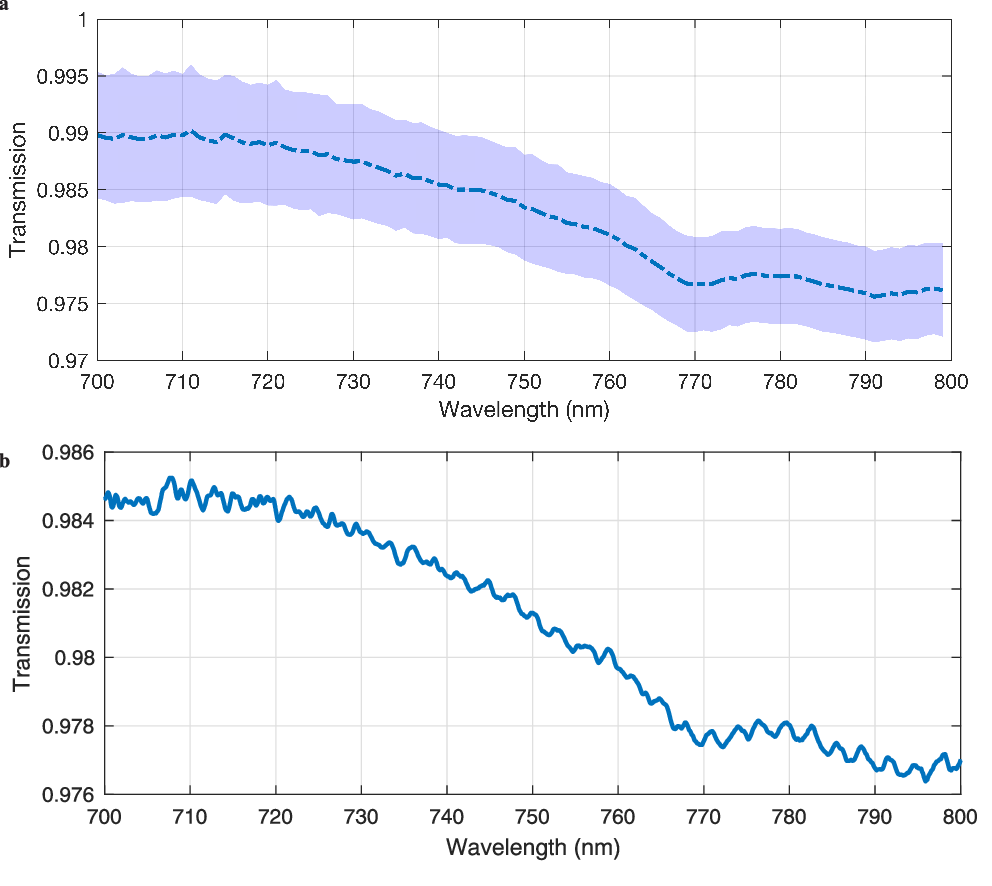}
\caption{\textbf{Characterization of a single CR of the PIC.} 
\textbf{a)} Mean value of the intensity transmission coefficient obtained for the CRs and the relative $2\sigma$ confidence interval. \textbf{b)} Example of the transmission curve for one CR. 
}
\label{fig:charcr_app}
\end{figure}

A spectral characterization of MMIs and CRs is carried out using a supercontinuum laser source combined with an optical spectrum analyzer, and averaging repeated measurements on 4 different chips. The obtained results for MMIs are reported in Fig. \ref{fig:charmmi_app}. At the working wavelength of the experiment, 750 nm, we obtain a normalized transmission coefficient $T=0.48\pm 0.02$ and a reflection coefficient $R=0.52\pm0.02$ (see Fig. \ref{fig:charmmi_app}(a).). The measured mean insertion loss is $0.8\pm0.4$ dB, as shown in Fig. \ref{fig:charmmi_app}(b). A positive insertion loss value is not physical. It is most likely due to issues with the straight waveguide used as reference (e.g. facets damages). Indeed, if we consider one of the chips, see Fig. \ref{fig:charmmi_app}(c), it is possible to observe that the insertion loss value is lower than 0, as expected. Since the fundamental parameters for the model of the PIC's real charactersitic are the normalized transmission and reflection coefficient values, we set the transmission coefficient $\beta$ of the MMI (see Section \ref{app:ni_model}) to $\beta=1$, as it just cancels out during the calculation.
Regarding the CR, the results of the characterization measurements are reported in Fig. \ref{fig:charcr_app}. The mean value of the transmission coefficient of the CR is calculated by performing the mean between the coefficient estimated for the four chips, whose transmission coefficients are measured by comparing three sequences of 50, 100 and 150 CRs. The measured value is $0.983\pm0.002$. An example of the transmission spectrum of a single CR from one PIC is reported in Fig. \ref{fig:charcr_app}(b), which is obtained from the transmission of a sequence of 150 CRs by a division for the total number of CRs.
\subsection{Phase-power relation}
\begin{figure}[]
\centering
  \includegraphics[width=\textwidth]{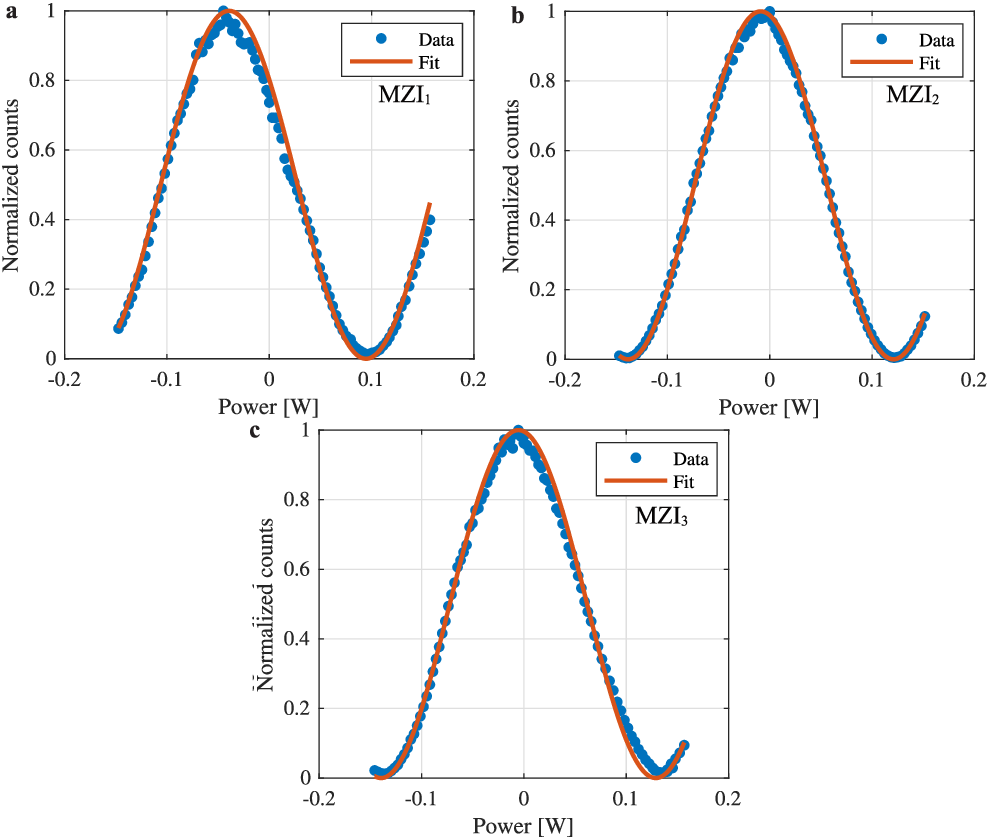}
\caption{\textbf{Characterization of the different MZIs in the preparation stage.} 
Characterization curves at 750 nm in TE polarization: \textbf{a)} $\text{MZI}_1$, \textbf{b)} $\text{MZI}_2$, \textbf{c)} $\text{MZI}_3$. The normalized counts are reported in blue, while the fit is reported in red.
}
\label{fig:charmzi_app}
\end{figure}

Prior to the experiment, all the MZIs on-chip have been characterized to determine the phase-power relation $\phi=\phi(W)$. This is retrieved by fitting the observed counts detected by a SPAD at the output ports of each MZI. The equations used, depending on the output port considered, are:
\begin{equation}
    N_{\text{out}1}=a\cos(b W+ d)^2 + c\,, \qquad
    N_{\text{out}2}=a\sin(b W+ d)^2 + c\,,
\end{equation}
where $a,b,c,d$ are fitting parameters. Consequently $\phi(W)=bW+d$.
The data collected and the relative fits are reported in Fig.~\ref{fig:charmzi_app}. 
It has to be noted that the waveguides of each arm of the MZIs are isolated from the surrounding substrate by means of trenches etched in the material. These result in a reduction of the heat diffusion coefficient helping to mitigate the thermal cross-talk between neighbouring PSs.
\begin{table}[!h]
    \centering
    \begin{tabular}{c||c|c|c|}
        &  $\text{MZI}_1$ & $\text{MZI}_2$ & $\text{MZI}_3$ \\
        \hline
         b[1/W]&$11.74\pm 0.05$& $12.11\pm 0.03$& $11.70\pm 0.05$  \\
         d     &$0.461\pm 0.004$& $0.106\pm 0.002$& $0.062\pm 0.004$ \\
    \end{tabular}
    \caption{\textbf{Fit parameters for the different MZIs.} Values of the parameters $b$ and $d$ obtained by the fitting operations.}
    \label{tab:resultfit_app}
\end{table}

\bibliography{sn-article}

\end{document}